# War: Origins and Effects

*How connectivity shapes the dynamics and development of the International System*

*Ingo Piepers*

*ingopiepers@gmail.com*

Version: 2 October 2014

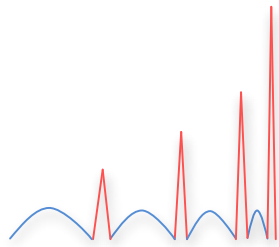

## Abstract

The International System is a self-organized system that shows emergent behavior. During the timeframe (1495–1945) covered in this study, a finite-time singularity and four accompanying accelerating log-periodic cycles shaped the dynamics of the International System. Each cycle began and ended with a systemic war. During their life span, these cycles show remarkable regularities in their dynamics. The accelerated growth of the connectivity of the International System's regulatory network – in combination with its anarchistic structure – produce and shape the war dynamics of the system. The accelerated growth of the International System's connectivity is fed by population growth and the need for social systems to fulfill basic requirements. The finite-time singularity and accompanying log-periodic oscillations were instrumental in the periodic reorganization of the International System's regularity network and contributed to a long-term process of social expansion and integration in Europe. The singularity dynamic produced a series of organizational innovations. At the critical time of the singularity (1939), the connectivity of the system reached a critical threshold and resulted in a critical transition that led to a fundamental reorganization of the International System: Europe transformed from an anarchistic system to a cooperative security community. This critical transition also marked the actual globalization of the International System. During the life span of the cycles, the war dynamics showed chaotic characteristics. These chaotic characteristics were temporarily eliminated during an exceptional period (1657–1763), during which the abnormal war dynamics affected the singularity dynamics of the International System. The stability of the International System and its development, the life spans of successive cycles and systemic wars, and the destructive power required to reorganize the International System were all closely related to the connectivity of the system, and the development of this parameter over time. Various early-warning signals can be identified and can most likely be used in the current International System, such as the signaling of upcoming systemic wars and a critical transition in the system. These findings have implications for the social sciences and historical research. The findings of this study enable the identification of the design principles for the regulatory network (governance) of the International System that ensure that the International System can reorganize in time without resorting to systemic war. Furthermore, this study sheds new light on the functioning of this category of complex systems.

*Key words: war, self-organization, singularity, accelerated network, log-periodic cycles, critical transition, non-linear dynamics, chaos, early-warning signals, social expansion, integration, governance, international system*

# 1. Introduction

In this study, I discuss the most significant findings of my research focusing on war dynamics and the development of the International System. I developed a new approach to study these dynamics that makes use of complex systems theory and network science. This approach resulted in some remarkable findings and a new theory or conceptual framework to better understand and analyze the dynamics of the International System. I have identified regularities – such as a singularity dynamic that is accompanied by accelerating log-oscillations – that determine the shape of war dynamics and the direction of developments in the International System to a large degree. In addition, I have identified the underlying mechanisms that generate these dynamics. Connectivity and the growth of connectivity seem to 'drive' these dynamics and to cause network effects that shape these dynamics.

The analysis developed herein shows that the singularity dynamic and the accompanying accelerating log-periodic oscillations of the International System are actually instrumental in the reorganization of the International System by introducing ever more intrusive organizational innovations. In 1939, i.e., the critical time of the singularity, the connectivity of the International System reached a critical threshold as a critical transition was triggered. This critical transition resulted in a cooperative security community in Europe, and marked the actual globalization of the International System. Until 1939, the International System had been unable to reorganize its regulatory network other than through systemic war: connectivity growth in an anarchistic system is a deadly dynamic.

This study shows that the International System is a deterministic system that self-organizes and shows emergent macro behavior. These emergent properties are the outcome of accelerated connectivity growth and multiple interactions at a micro level.

This study may potentially have a great impact on the social sciences: it shows that to fully comprehend the dynamics of the International System, it does not suffice to analyze only the dynamics *on* the network of the International System. It is also important to analyze the dynamics *of* the underlying network and the interplay between both levels.

It is safe – and wise – to assume that the current International System (post-1939 singularity) potentially exhibits the same destructive singularity dynamics: the International System remains an anarchistic system, and the connectivity of the system continues to grow. The deadly combination still exists.

Based on these new insights, it is now possible to design a regulatory network for the International System that does not eventually collapse, resulting in a destructive systemic War. Designing such a network requires international cooperation, which history shows is difficult to achieve. I hope that these insights will provide new energy and direction to research into the dynamics of the International System, will help design and organize a set of early-warning signals, and will provide some initial guidelines for the development of a robust and war-free International System: we should focus our energy and resources on problems that collectively as a species, we must solve – such as poverty and climate change – to improve our chances for survival. Thus, we must understand the trap that the system has set for us.

In section 2, I explain the methodology I use and provide information about the dataset that was crucial in identifying the emergent properties of the International System. In section 3, I introduce two perspectives on international politics: the 'traditional' perspective, and the perspective I have developed that is based on insights from complex systems theory and network science. This section will introduce the reader to some definitions, and the different approaches of both perspectives.

Section 4, "Acceleration, saturation, collapse, and systemic war," introduces regulatory networks that require accelerated growth to achieve their functional objectives. However, acceleration at the rate that is required is unsustainable: at a certain stage, the network becomes saturated and collapses. I argue that the International System is such an accelerated regulatory network that periodically – and with remarkable regularity – collapses. I define these collapses – wars on a system-wide scale with specific characteristics – as systemic wars. Four of these wars are identified. These systemic wars typically result in new organizational innovations that ensure a new relatively stable period of growth.

Section 5 discusses finite-time singularities that are accompanied by accelerating log-periodic oscillations, including the finite-time singularity that can be identified in the war dynamics of the International System.

Connectivity and network effects are the subjects of section 6, which discusses a model developed by Watts and its applications for this study.

The analysis of the war data shows that during a specific period during the life span of the second cycle – a cycle is a relatively stable period in between two successive systemic wars – the war dynamics of non-systemic wars showed a temporarily abnormal dynamic. This abnormal dynamic, in combination with other clues, may point to the *chaotic* characteristics of non-systemic war dynamics. This subject is discussed in section 7, titled "War dynamics with chaotic and periodic characteristics."

In section 8, I define four different types of war that can be distinguished from one another and discuss the development of certain characteristics of the International System, such as its stability and resilience over time.

Early-warning signals are the subject of section 9. With the help of plainly identified regularities and a clear direction for the development of the International System toward more cooperation, it is possible to identify early-warning signals that may possibly be warning signals for wars – particularly for upcoming systemic wars – in the current International System.

In section 10, 'Recapitulation and implications,' I elaborate on the findings and potential implications of this study.

## 2. Methodology and data

*Introduction.* In this section, I explain the methodology of this study, introduce certain definitions, and provide information about the data I used. In this study, I describe a new theory – a new 'framework' – to explain the origins and the effects of the 'war' dynamics of the International System. The war dynamics during the 1495–1945 period are studied, and Europe is at the heart of these dynamics. This new approach is based on theories and insights from complex systems theory and network science as well as on recent results from research in other scientific disciplines.

Self-organized systems can show macroscopic 'emergent' behavior that results from multiple interactions on a micro level. This macroscopic behavior has its own logic that cannot be derived from these micro-interactions. Can such macroscopic emergent behavior – patterns, regularities, and their mechanisms – be identified in the dynamics of the International System? This study shows that the International System does show emergent behavior and that the underlying mechanisms can be identified with the help of complex systems theory and network science.

The focus of my research, above all, is on war – and on war dynamics, in particular – and on its relation with the development of the International System. I opt for a long-term perspective, hoping that regularities – if they exist – and the underlying mechanisms that cause these regularities may then be identified more easily. For example, phenomena such as phase transitions can best be identified at an aggregate level (30).

*Levy's dataset and assumptions.* For this study, I mainly used Levy's dataset (34). This dataset is considered as accurate and complete as such a dataset can be; Levy's dataset is available in 'supporting information.'

Levy writes: "The Great Power framework shares the basic assumptions of the realist paradigm of international politics but focuses explicitly on the small number of leading actors in the system. It is assumed that in any anarchic international system there exists a hierarchy of actors determined on the basis of power. In the modern system since 1500, the dominant actors have been dynastic/territorial states and nation-states; in the international system of ancient Greece and Renaissance Italy the dominant actors were city-states. The more powerful states – the Great Powers – determine the structure, major processes, and general evolution of the system. Therefore, the actions and interactions of Great Powers are of primary interest. Secondary states and other actors have an impact on the system largely to the extent that they affect the behavior of the Great Powers. This hierarchy of actors is intimately related to the hierarchy of issues dominated by military security. It is assumed that issues overlap and that the currency of military power is applicable to and effective in the resolution of other issues. The concept of a Great Power system is based on the traditional assumption, shared by realists since Thucydides, that world politics is dominated by security issues and the struggle for power.

The priority of military security derives from the perception of a high-threat environment, which in turn derives primarily from the anarchic structure of the International System. In this context, it is the Great Powers, because of their military capability and ability to project it, which generally can do the most to affect the national interests of others and are therefore perceived as the most serious security threats. Consequently, the Powers direct their primary attention toward each other. A relatively high proportion of their alliance commitments and war behavior is with each other, and they tend to perceive international relations as largely dependent upon and revolving around their own interrelationships. The general level of interactions among the Great Powers tends to be higher than for other states, whose interests are narrower and who interact primarily in more restricted regional settings. Thus the Great Powers constitute an interdependent system of power and security relations, which will be called "the Great Power system" (34, p8-9).

Levy defines a "Great Power as a state that plays a major role in international politics with respect to security-related issues. The Great Powers can be differentiated from other states by their military power, their interests, their behavior in general interactions with other powers, other Powers' perception of them, and some formal criteria" (34, p16).

*Methodology*. I follow an iterative approach: based on an analysis of the dataset and on a 'complex systems theory and network perspective,' I formulated tentative descriptions and hypotheses with respect to the workings of the International System. Based on some of these assumptions, it was possible to make some tentative predictions that could be falsified to a certain degree. Similar research into the dynamics of cities, financial markets, ecosystems, and networks was helpful to me in structuring this process.

Conceptual consistency was also an important 'guideline' to assess various ideas and predictions. Furthermore, in some cases, the research and assessments of historians was helpful to produce a better 'context,' and to find useful clues for 'underlying' mechanisms. Historical interpretation was useful, for example, in explaining the abnormal war dynamics during an exceptional period (1657 -1763) during the span of the second cycle.

It is impossible to scientifically prove some of the claims I make in this study. The outcome remains in some respects speculative and further research to validate (falsify) some of the assumptions I made and hypotheses I used and propose is necessary. It is frequently 'circumstantial' evidence that 'proves' the plausibility of certain hypotheses. However, the theory seems consistent and is actually quite simple. In addition, its predictive power is significant.

### 3. <u>Two Perspectives</u>

*Introduction*. This study demonstrates that social and historical analysis is focused above all on what I call the dynamics *on* the network of the International System, not on the dynamics *of* the underlying network. I will show that these concepts should be studied in conjunction, including the interplay between both levels.

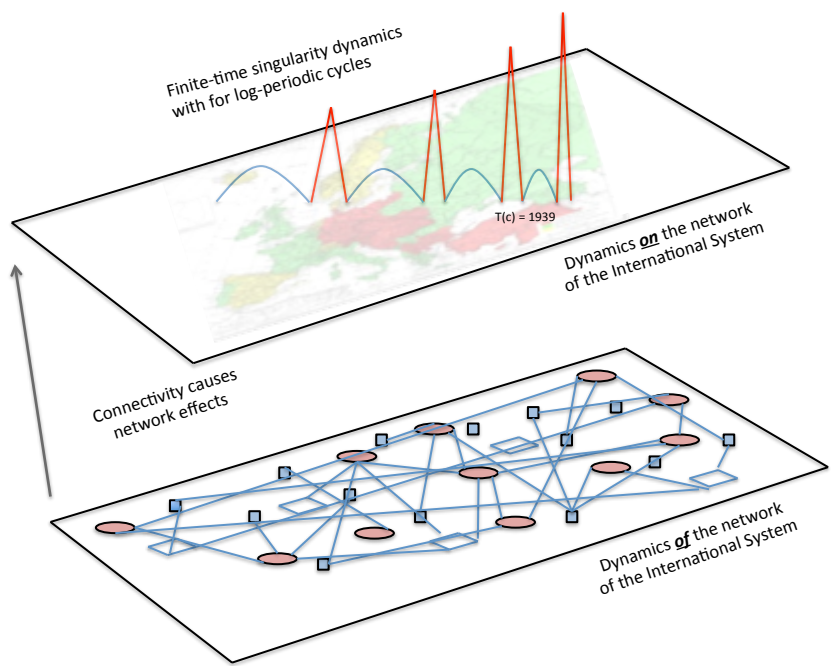

*Figure 1. This figure shows schematically the two levels of dynamics that can – and must – be distinguished: the dynamics **on** the network of the International System that social scientists and historians typically focus on and the dynamics **of** the underlying network. The interplay between both levels also must be considered. This study shows the strong shaping effects of the (growth of) the connectivity of the underlying network of the International System on the dynamics on the network.*

In this section, I describe the typical 'one-level' approach of political scientists and historians. The purpose of this section is to introduce certain terminology and to become familiar with the subject. In particular, I cite Gilpin because his focus is on war and change, as well. In fact, in 'War and Change in World Politics' (23), Gilpin addresses "the role of war in the process of international change." In this section, I will cite certain key elements of his approach.

*A social science perspective*. Gilpin defines the International System, "as an aggregation of diverse entities united by regular interaction according to a form of control" (23, p26). This definition is generally accepted. For example, Holsti defines the International System in similar terms: "any collection of independent political entities – tribes, city-states, nations, or empires – that interact with considerable frequency and according to regulated processes" (28).
Gilpin notices that the balance of power among the actors of the system change as a result of economic, technological, and other developments. According to Gilpin, the cause and the consequence of international political change are "the differential growth of power in the system."
The nature of the International System determines whose interests are being served by the functioning of the system (23, p10). Actors in the International System are assumed to behave as though they were guided by a set of cost/benefit calculations (23, p11). "An international system is in a state of equilibrium if the more powerful states in the system are satisfied with the existing territorial, political, and economic arrangements" (23, p11).

Gilpin argues that "an international system or order exists in a condition of homeostatic or dynamic equilibrium.... it is not completely at rest; changes at the level of interstate interactions are constantly taking place. In general, however, the conflicts, alliances, and diplomatic interactions among the actors in the system tend to preserve the defining characteristics of the system" (23, p12).

However, "in time, the differential growth and power of the various states in the system causes a systemic disequilibrium, and hegemonic war as a result." Gilpin writes: "This disjuncture (a systemic disequilibrium) within the existing international system involving the potential benefits and losses to particular powerful actors from a change in the system leads to a crisis in the international system. Although resolution of a crisis through peaceful adjustment of the systematic disequilibrium is possible, the principal mechanism of change throughout history has been war, or what we shall call hegemonic war (i.e., a war that determines which state or states will be dominant and will govern the system). The peace settlement following such a hegemonic struggle reorders the political, territorial, and other bases of the system. Thus the cycle of change is completed in that hegemonic war and the peace settlement create a new status quo and equilibrium reflecting the redistribution of power in the system and the other components of the system" (23, p15).

According to Gilpin, an international system has three primary aspects: (1) "diverse entities," which may be processes, structures, actors, or attributes of actors; (2) "regular interaction," which may vary on a continuum from infrequent contacts to intense interdependence of states; and (3) some "form of control" that regulates behavior and may range from informal rules of the system to formal institutions.

"A view prevalent among many scholars of political science is that the essence of international relations is precisely the absence of control." "International politics are said to take place in a condition of anarchy."

Gilpin's argument is "that the relationship among states has a high degree of order and that although the international system is one of anarchy, the system does exercise an element of control over the behavior of states." "Control over the governance of the international system is a function of three factors. In the first place, governance of the system rests on the distribution of power among political coalitions" (23, p28).

"The second component in the governance of the international system is the hierarchy of prestige among states." Ultimately, this hierarchy, according to Gilpin, "rests on economic and military power" (23, p30).

The third component "is a set of rights and rules that govern, or at least influence the interactions among states" (23, p34). These "rest on common values and interests and are generated by cooperative action among states" (23, p35).

However, the fact that the International System is an anarchistic system has its consequences: it can 'activate' what in political science is called a 'security dilemma.' In an anarchistic system, states are – and feel responsible for – their own security. Security can be achieved by respecting ('peaceful') rules of the system – by respecting international law and incentivizing other states to do the same – but also by military capabilities and participating in alliances. However, and this constitutes the security dilemma, what is security for one state – military capabilities and alliances – is (potentially) insecurity for other states. The security dilemma can cause positive feedback loops, i.e., 'arms races' between states and alliances.

Social systems in which wars and large-scale military violence have become rare or even unthinkable are called 'security communities.' Deutsch defines a security community as "a group of people 'believing' that they have come to agreement on at least this point: that common social problems must and can be resolved by processes of peaceful change" (18). The United States, and more recently the European Union, qualifies as a security community.

Gilpin also distinguishes three types of international change. For the purpose of this study, in particular, a "change in the form of control or governance of an international system" is of interest. These types of changes are called "systemic changes" (23, p40). Gilpin observes as follows: "The most frequently observed types of changes are continuous incremental adjustments within the framework of the existing system" (23, p45).

Gilpin and most historians and social scientists reject an overly deterministic type of interpretation of political change. "Although it is certainly possible to identify crises, disequilibria, and incompatible elements in a political system – and to identify a disjuncture between governance of the system and the underlying distribution of power, in particular – it is most certainly impossible to predict the outcome. In the social sciences, we do not have (and will likely never have) a predictive theory of social change in any sphere. Although we observe international crises and corresponding responses in the behavior of states, it cannot be known in advance if there will be an eventual return to equilibrium or a change in the nature of the system."

This study demonstrates that contrary to the assumption of many social scientists (19)(22)(23)(24)(54), a predictive theory of social change in the International System can be formulated, which allows 'returns to equilibrium or a change in the nature of the system" to actually be known in advance.

*A complex systems theory and network perspective*. I have defined the International System from a complex systems theory and network perspective (3)(5)(6)(7)(32)(38)(48). Thus, I define the International System as a complex network of international issues and actors that are connected by different types of interactions. This network consists of actors (stakeholders) that are linked or connected to one another and to various 'international issues.' An issue is a question – not necessarily a problem – that concerns more than one 'international' actor. Issues have stakeholders. Decision makers representing a state, politicians, non-governmental organizations, the UN, policymakers, and also civilians can be stakeholders with respect to particular issues. For actors, three types of interaction are available: cooperation, competition, and conflict.

The International System is a subsystem of the Global System. From this perspective, 'states' – and their issues and their interactions – constitute a subsystem of the International System. I consider the Great Power System a subsystem of the International (and Global) System(s) and to 'represent' the International System for the purposes of this study.

The International System is a continuously evolving – and growing – multilayered, and hierarchically organized (47) network of issues and actors. Various types of 'social systems' – such as states – are 'components' or building blocks of this network.

As discussed above, the International System lacks an overall legitimate and formalized governance structure: thus, it is anarchistic. However, control that is formalized in institutions and through informal mechanisms is an essential requirement for a network such as the International System to enable its components – i.e., its members – to function.

In this study, I regularly refer to the regulatory network of the International System. This particular network does not consist solely of the formal governance structures (institutions and rules) of the International System. The regulatory network consists of much more, such as informal rules, informal communication, alliances, etc.

The International System also qualifies as a self-organized system that shows emergent behavior at the macro level. The results from this study make this emergent macro behavior 'visible.' The emergent dynamics 'originate' from the regular interactions of its components, such as states, in this context. Typically, this emergent macroscopic behavior cannot be explained by the micro-level behavior of its components: it develops its own 'indivisible' dynamic and logic. From this perspective, conflict interactions, must be understood as an attempt to fulfill basic requirements and/or to achieve internal and external rebalancing.

*Basic requirements.* Another assumption that I make is that social systems must fulfill four interdependent categories of basic requirements to function and survive; see Table 1 (13)(38) (39) below.

| Basic requirements of social systems | |
|---|---|
| **Basic requirements** | **Subsystem** |
| (1) Energy, necessities of life, and (2) wealth. | Economic system |
| (1) Internal and external security and (2) the potential to influence the behavior of individuals and other (sub)systems. | Threat system |
| (1) Individual and collective identity and (2) the development of individual and collective identities. | Value system, (religion, culture) |
| (1) Internal and external consistency and balancing, (2) direction for the development of the system, (3) legitimacy/acceptance of the (political) leadership of the system, and (4) the potential to control the environment of the social system. | Integrative system |

*Table 1. This table provides an overview of four basic requirements that social systems must ultimately fulfill to survive. The integrative system provides control and integration. Control and direction of social systems, including the International System, require a certain degree of predictability and the availability of mean values and measures to react to changes.*

Each basic requirement delivers a certain set of 'services' to the social system it supports and of which it is an integral part. The International System should also provide certain 'services' to its 'members' to help them meet their basic requirements. The ability of the International System to deliver these services depends on various factors. If the International System loses its effectiveness, there are consequences for its already (by definition) limited 'legitimacy.'

The four requirements are closely related; they overlap and require a certain level of consistency and internal and external balancing. Consistency and balancing of each basic requirement is necessary not only in relation to the other (three) basic requirements of the same system but also in relation to the other basic requirements of other social systems.

To achieve consistency, three types of interaction are available: cooperation, competition, and conflict. The type of interaction that actors choose to adopt under certain circumstances (to achieve the required output of the basic requirement and to achieve consistency and balancing) is a deliberate choice made by the decision makers of the social system and more often is likely the outcome of 'internal' discussions and balancing of interests.

The basic requirements, their specific conditions, the need for optimization, and the use of economies of scale result in an organization of the International System with specific functional characteristics. Fulfilling and balancing various requirements necessitate information integration.

The International System is not a fully developed social system; for example, it lacks legitimate control and cannot always provide for or guarantee the security of its 'members.'

Fulfilling basic requirements and achieving and maintaining their consistency in an ever growing and evolving system are complicated; equilibrium is a relative term, and at best, a temporary condition, particularly when resources are scarce in an anarchistic setting.

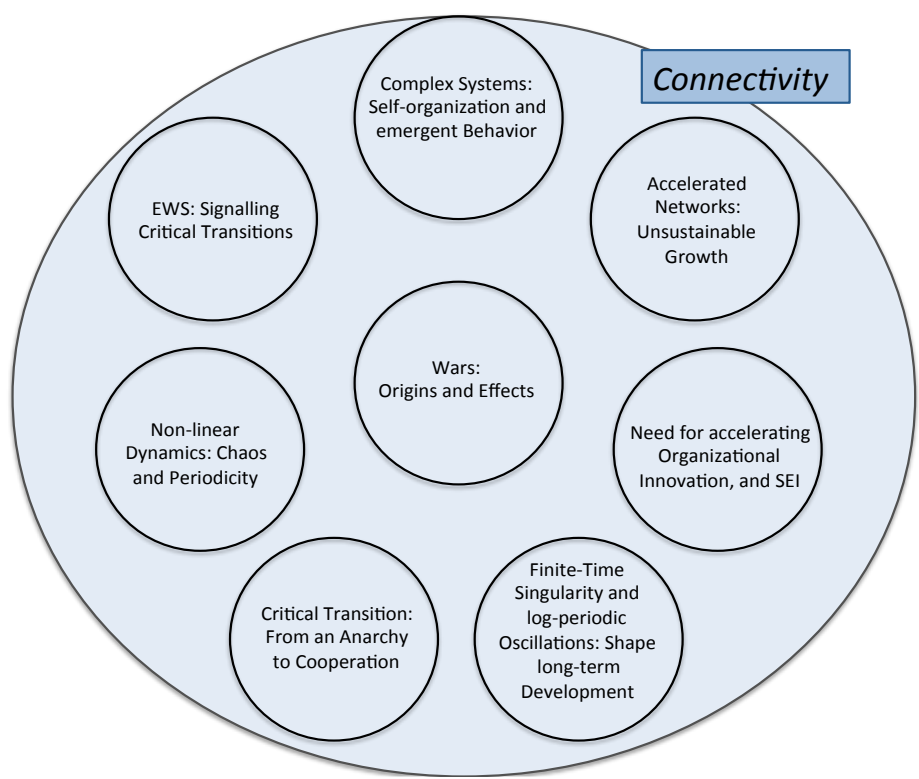

*Figure 2. This figure shows the different 'points of view' of this study.*

## 4. Acceleration, saturation, collapse, and systemic war

*Introduction.* In this section, I discuss findings in research related to cities, attributes of these cities, and how these attributes develop with population size (8). There is an interesting relationship between the pace of life in cities and their size. Next, I will discuss 'accelerated networks' and their typical dynamics. I assume that the regulatory network of the International System is actually just this type of network. However, it is problematic – as will be explained – that these types of networks (including the International System) have certain limits.

*Cities.* Bettencourt et al. (9)(10)(11) and Arbesman et al. (2) find that the scaling exponents for urban properties and their implications for growth can be classified into three groups. When the scaling exponent < 1, these systems and networks typically optimize for efficiency. The attributes belonging to this category show economies of scale. The larger the size of the population, the more efficient its 'operation': thus, the number of gasoline stations in a city decreases with population size.
However, for the "creation of information, wealth, and resources," the scaling exponent > 1, which implies superlinear growth. Superlinear growth has limits, however, and cannot be sustained indefinitely, with a 'boom/collapse' dynamic resulting (9).
Exponential (that is 'accelerated') growth cannot be maintained and results at a certain stage in discontinuities and collapse. Bettencourt et al. suggest that some of their findings "very likely generalize to other social organizations, such as corporations and businesses, potentially explaining why continuous growth necessitates an accelerating treadmill of dynamical cycles of innovation." The effect of such cycles is a "reset of the singularity and postponement of instability and subsequent collapse" (9).

In this section I explain that such an 'accelerated growth' dynamic could well be present in the dynamics of the International System, which I show is actually the case in sections 5 and 6 below.

Schlaeper et al. (45) and Arbesman (2) argue that "increasing social connectivity underlies the superlinear scaling of certain socioeconomic quantities with city size." Schlaeper et al. assume that "network densification facilitates interaction-based spreading processes as cities get bigger." In other words, connectivity drives the 'pace' of social systems, and connectivity growth creates a speeding-up effect.

*Accelerated networks and their characteristics*. There are different types of networks, each with specific characteristics. For this study, the distinction between 'simple connection networks' and regulatory networks is important. Regulatory networks can be found in biology (e.g., the network of regulatory proteins that controls gene expressions in bacteria), engineering (e.g., computer operating systems), and society. I assume that the International System is also such a regulatory network: the 'components' constituting the International System must be able to operate in a globally responsive manner. In particular, Great Powers require global responsiveness to maintain their position and to serve their interests; global responsiveness requires the availability and integration of information. The regulatory network is in fact a subsystem of the International System, consisting of formal institutions, formal and informal rules, information systems, etc. (see section 3).

Mattick et al. (35) explain that – as opposed to simple connection networks – regulatory networks cannot grow unconstrained: "Regulatory networks are accelerating networks that must be able to operate in a globally responsive way."

Mattick et al. argue that in particular, functionally organized systems "whose operation is reliant on the integrated activity of any or all its component parts" require sufficient connections to ensure their operation: in these types of systems, "the number of informative connections per node must increase with the size of the network." This requirement indicates that the total number of connections between nodes – states and issues in the context of this study – must scale faster than they would be able to linearly with the number of nodes. For that reason, such networks are denoted "accelerated networks."

"These accelerating connection requirements impose an upper limit on the functional complexity that integrated systems can attain." Mattick et al. explain: "the number of connections must scale quadratically, otherwise global connectivity will decline." "This in turn indicates that the size and complexity of such systems must sooner or later reach a limit where the number of possible connections becomes saturated or where the accelerating proportional cost of these connections becomes prohibitive." At a certain point, the connectivity of the network reaches a critical threshold.

Various characteristics of the Global System and the International System place high demands on the regulatory network of the International System, such as the following: (1) the functional organization of the system (resulting from the 'organization' of basic requirements and the need to balance these requirements internally and externally), (2) the organization of the International System into more or less autonomous states, and (3) the anarchistic 'setting' of the system. This anarchistic characteristic requires that states and particularly Great Powers must always 'monitor' developments in the system that are or may become a threat.

Because the regulatory network of the International System is an accelerated network, it cannot grow in an unrestricted manner; growth will sooner or later result in saturation of the regulatory network and its subsequent collapse.

Mattick et al. explain that global responsiveness "imposes an upper size limit on the complexity of integrated systems due to the costs incurred by the need for an increased number of connections and levels of regulation." This limitation indicates that these types of networks cannot grow in an unlimited manner because they must "rapidly integrate information from, or globally respond to the current state of their nodes."

I argue that the accelerating connection requirements of growing regulatory networks also have implications for the functioning of the International System: the size and complexity of the regulatory network – and as a consequence of the International System – for that reason (sooner or later) reach a limit "where the number of possible connections becomes saturated or where the accelerating proportional cost of these connections becomes prohibitive." Growth in the connectivity of the International System has a limit and comes at a price.

The limit that Mattick et al. describe "can be breached by a reduction in connectivity, which however reduces the functional integration of the network, leading to fragmentation, as is for example observed, in the transition of social networks from small communities to cities." "However, if integration of node activity is absolutely required" – Mattick et al. explain – "for the operation of the system or for its competitive survival, the functional complexity of the system can only be increased beyond the existing limit by increasing the number of connections."

A similar dynamic can be observed in the dynamics of the International System. I argue that the International System is not able to sustain the accelerating growth of its regulatory network, which periodically results in what I define as systemic wars. These systemic wars differ fundamentally from other wars that I call non-systemic wars and are 'outliers' not only in their function but also in their number. Four out of 119 wars in Levy's dataset (34) qualify as systemic wars.

Systemic wars (1) achieve a reduction in the connectivity of issues and (2) reorganize the system by means of war. Systemic wars always result in the introduction of new organizational mechanisms – innovations – to improve the organization of the International System and to achieve better cooperation between its members.

A systemic war results from the unsustainable growth of the connectivity of the regulatory system. However, systemic wars typically 'offer' a new solution: every systemic war resulted in a reorganization of the governance structure, enabling more and better information processing and global coordination.

From a dynamic systems perspective, these systemic wars can be defined as periodic resets of the parameters of the system, allowing for the next phase of growth. In 1939, as I discuss more fully below, the possibilities to innovate – to reset the parameters of the system within the context of the current anarchistic structure – had reached their limits, resulting in a critical transition of the system itself. A closer examination shows that the Second World War was a singularity of the International System. This singularity was preceded by four accelerating log-periodic oscillations consisting of three systemic wars. The singularity and accelerating log-periodic oscillations show a remarkable regularity, as I explain in the next section. The singularity dynamics and log-periodic oscillations constitute accelerating cycles of the organizational innovation of the regulatory network.

Mattick et al. describe a more fundamental solution for the unsustainable growth of regulatory networks as follows: "When connection limits (the critical threshold) cannot be raised, or functional components cannot directly communicate (anymore) with each other, the alternative is to introduce dedicated hierarchies, called management in organizations, control systems in engineering, and regulation in biology."

This description of the development path of the International System is accurate: the outcome of the singularity dynamic is the introduction of dedicated hierarchies into the European System in combination with the establishment of a security community.

The connectivity growth of 'Europe' – the outcome of a long-term historical process – could not be sustained in an anarchistic system. For a certain period of time (until 1939), the regulatory network problem could be 'fixed' with periodic organizational innovations within the context of the anarchistic system. These types of solutions reached their limit in 1939, causing a systemic war and resulting in what now seems to be an unavoidable 'critical transition.'

The same prerequisites that allowed for temporary improvements (the first three systemic wars) of the regulatory network set up the International System for a large-scale collapse (the fourth systemic war, i.e., the Second World War). These types of dynamics are also observed in ecosystems (44).

In the end, cooperation proved to be stronger than conflict (anarchy) as the preferred type of interaction in Europe. The singularity dynamic, beginning in 1495, was instrumental in the process of social expansion and integration in Europe. Under what conditions would Europe have fragmented as a consequence of the force of the singularity dynamic and not have achieved this favorable outcome? Was 'success' – albeit at extremely high costs – inevitable? These are both interesting and relevant questions for the development of the current International System.

Mattick et al. argue that "these hierarchies (referring to hierarchical organizational structures) have their own costs. Each level of regulatory hierarchy introduces time delays, increases noise and stochastic errors." "These shortcomings increase with greater levels of regulation and with network size, limiting system coherence and ultimately imposing (new) upper limits on the size and functional complexity that such systems can attain." This dynamic can be observed in the 'European Project' as well, and can cause fragmentation as a result. Nonetheless, in this study, I will not elaborate further on this subject.

## 5. The finite-time singularity and log-periodic oscillations of the International System

*Introduction*. The previous section discussed accelerated regulatory networks and their inherent limitations. Mattick et al. argued: "Systems that require integral organization to function in a competitive environment are dependent on, and ultimately constrained by, their accelerating regulatory architecture. Thus, connectivity and the proportion of the system devoted to regulation must scale faster than function in organized complex systems." Consequently, "periodic constrains must be relieved." Typically, "accelerating networks show quasi-stationary phases of growth in their complexity and capability, asymptotically approaching maxima until the ceiling is lifted."

The cycles between two consecutive systemic wars that appear at the beginning and the end of these cycles are in fact such quasi-stationary phases Mattick et al. describe and demonstrate growth in the complexity of the International System until a new point of regulatory saturation is reached, with a systemic war as its consequence.

In this section, I closely examine other singularities, and I find clues for the 'driver' of the singularity dynamic in the International System. I then identify the organizational innovations that were introduced by systemic wars.

*Singularities*. Various researchers (29)(31)(49)(50) have studied systems that show faster than exponential growth. Kapitza (31) has focused on the growth of world population, and Sornette et al. (49) have studied economic and financial indices, in addition to world population growth.

As explained in the previous section, growth that is faster than exponential is unsustainable. In this section, I discuss so-called singularities and the dynamics they present and generate: I argue in this study that a singularity with a critical time of 1939 – accompanied by four log-periodic cycles – shaped the war dynamics of the International System for the long term since 1495.

Singularities "are mathematical concepts of natural phenomena: they are not present in reality but foreshadow an important – and unavoidable – transition or change of regime." In this context, "they must be interpreted as a kind of 'critical point' signaling a fundamental and abrupt change of regime similar to what occurs in phase-transitions" (30).

The researchers discussed above have determined that the human population and its economic output have grown faster than exponentially. Sornette et al. argue that a power law is an adequate model for this super-exponential growth of world population, world GDP, and various financial indices (30). As we have previously discussed, these growth rates are unsustainable, and in mathematical terminology, 'reach' an asymptote at a finite time.

This result indicates that "the acceleration of the growth rate contains endogenously its own limit in the shape of a finite-time singularity to be interpreted as a transition to a qualitatively new behavior." This phenomenon is the same as that described by Mattick in the previous section but is now approached from a somewhat different angle.

Consistent with the assumptions made by other researchers in other disciplines (45), I assume that the growth of connectivity is the driver of the International System and produces the singularity dynamics that are accompanied by accelerated oscillations.

I assume that the following causality is at play: population growth results in the connectivity growth of social systems. The growth in connectivity is related to the fulfillment of various requirements of individuals and social systems and is ultimately related to the ability of individuals and social systems to survive. This dynamic, in combination with economies of scale that can be achieved by cooperation, sets in motion a long-term process of social expansion and integration. These 'variables' create a self-reinforcing – i.e., positive feedback – dynamic.

(Global) integration and coordination require that the regulatory network of the International System grow at a super-exponential rate. Such a growth rate cannot be sustained and results in saturation of the network and in periodic systemic wars in an anarchistic context. Systemic wars provide organizational solutions – innovations – to improve the system's capabilities for cooperation. The rate of growth of connectivity has an accelerating effect, and relatively stable periods (in between systemic wars) become progressively shorter. A reset of parameters (the effect of systemic wars, at that stage) reaches its limits insofar as the provision of solutions is concerned. The connectivity of the network reaches a critical threshold. The singularity of the International System arrived at its critical time in 1939. The Second World War brought about a critical transition of the European International (sub)system; thus, the establishment of a security community in Europe also marked the actual globalization of the (war) dynamics of the Global System.

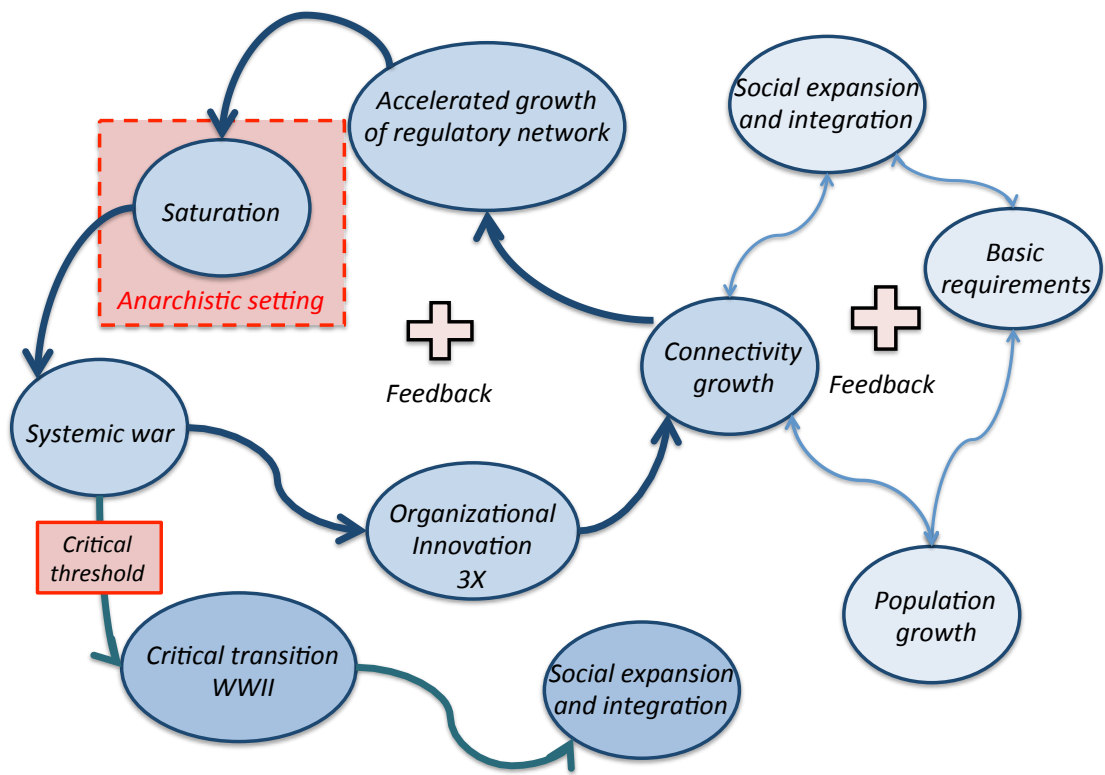

*Figure 3. This figure schematically presents the causality in the process described in this study. Various positive feedbacks are 'at work' and ultimately generate the singularity dynamics, and the accompanying accelerating log-periodic cycles.*

*Organizational innovations.* A closer examination reveals that four wars – I define them as systemic wars – led to the introduction of new 'innovative' organizing principles: the Thirty Years' War (1618–1648), the French Revolutionary and Napoleonic Wars (1792–1815), the First World War (1914–1918) and the Second World War (1939–1945). These wars are also identified by historians as wars with an enduring impact on the dynamics of the system that provided for relative stability during the next cycle. These wars constitute the singularity dynamic, accompanied by accelerating log-periodic oscillations, beginning in 1495. This dynamic shaped the International – and Global – System to a high degree.

The fourth systemic war, the Second World War, was in fact a critical transition of the system, fundamentally changing the structure and dynamics of the European continent. Until 1939, Europe was the core – the 'motor' – of the war dynamics of the International System. The critical time of 1939 also marked the 'globalization' of the International System. As a consequence of this critical transition, Europe 'transformed' from an anarchistic system into a cooperative security community.

In the table below, I specify the organizing principles that were introduced by the first three systemic wars.

| Accelerating cycles of organizational innovations | | | |
|---|---|---|---|
| **Systemic War** | **Time** | **Organizational innovation** | **Remarks** |
| Thirty Years' War | 1618–1648 | Peace of Westphalia: Sovereignty and principle of 'balance of power.' | |
| French Revolutionary and Napoleonic Wars | 1792–1815 | Concert of Europe: Periodic consultation. | |
| First World War | 1914–1918 | Versailles Peace Treaty: League of Nations, consultations, and limitations on 'behavior' of states. | Due to various (network) conditions, the aims of this reorganization cannot be achieved within the prevailing 'context' and network at the moment of its development. |
| Second Word War | 1939–1945 | Embedding cooperation in European structures, and establishing the United Nations (global reach). | This systemic war constitutes a critical transition: 1939 is the critical time of the singularity. |

*Table 2. Through each successive systemic war, new organizational innovations were introduced into the International System. All these innovations improved cooperation. The first three innovations were introduced within the anarchistic setting of the current system and not did change its fundamental structure and dynamics. The Second World War can be identified as the critical transition of the International System at the critical time of 1939. These systemic wars constitute a robust singularity dynamic accompanied by accelerating log-periodic oscillations.*

*Log-periodicity of oscillations*. The log periodic oscillations relate to this singularity and shape the war dynamics and the development of the International System. Each systemic war 'creates' a new quasi-equilibrium, i.e., a cycle with a life span. The shortening is visible in the table below. The shortening of the life span develops according to a simple mathematical formula.

| Development of life spans of oscillations | | | |
|---|---|---|---|
| Systemic War | | Time | t(c) – t |
| Second World War | t(c) = t-critical | 1939 | 0 |
| First World War | t(1) | 1914 | 25 |
| French Revolutionary and Napoleonic Wars | t(2) | 1792 | 147 |
| Thirty Years' War | t(3) | 1618 | 321 |

*Table 3. This table shows the calculations for the singularity-dynamic and accompanying accelerating log-periodic oscillations. The singularity 'behaves' remarkably consistently.*

The Life Span (LS) of successive oscillations can be calculated as follows: **LS(t) = 19.6e^(0.936 t) with R2 = 0.9918**. Because these oscillations are periodic in the logarithm of the variable (K(c) – K) / K(c), we refer to them as 'log-periodic.'

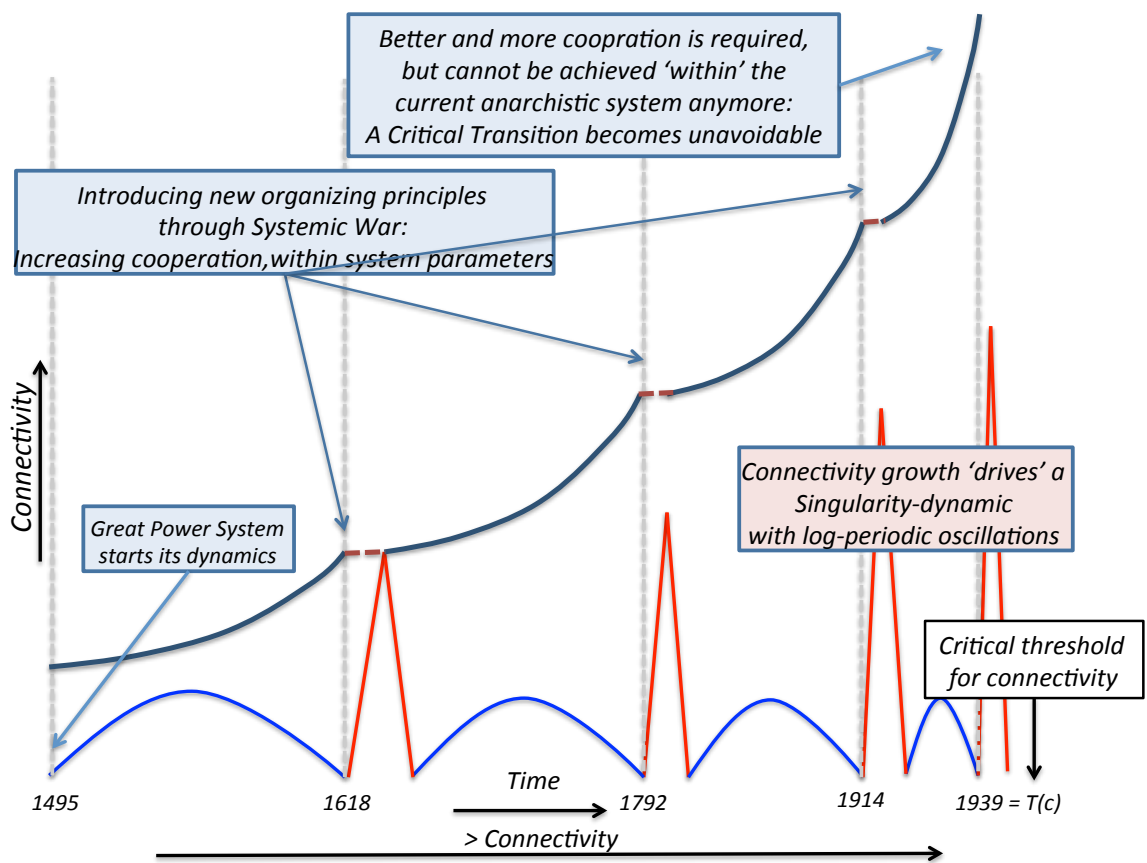

*Figure 4. This figure is a schematic image of the singularity dynamic and accelerating log-periodic oscillations that shaped the dynamics and development of the International System. The International System begins the dynamic at the start of the Great Power System, in approximately 1495. Successive systemic wars (in red) and four quasi-stable cycles can be identified. The life spans of the first three systemic wars, and of the successive cycles, develop regularly. Through the first three systemic wars, organizational innovations were introduced within the anarchistic setting of the International System. The fourth systemic war – the Second World War – constitutes a critical transition (t(c) = 1939), resulting in a fundamental change in the structure and dynamics of the European International System: the transformation from an anarchistic system to a cooperative security community. At the time of the singularity, the connectivity of the International System reached a critical threshold. In the preceding years – during the life span of the fourth cycle – the International System was unable to (re)create a certain balance, which is, in retrospect, an indication of the saturation level of the regulatory network. In 1939, the historian Carr published a study on international relations with the striking title 'The Twenty Years' Crisis, 1919-1939' (14). The Second World War also marked the actual 'globalization' of the International System. The intense rivalry between 'East' and 'West,' the United States and the Soviet Union, 'froze' the war dynamics temporarily in the system, providing 'time' for 'Europe' to embed new structures and rules within its system. This study shows that connectivity growth drives the war singularity dynamics of the International System.*

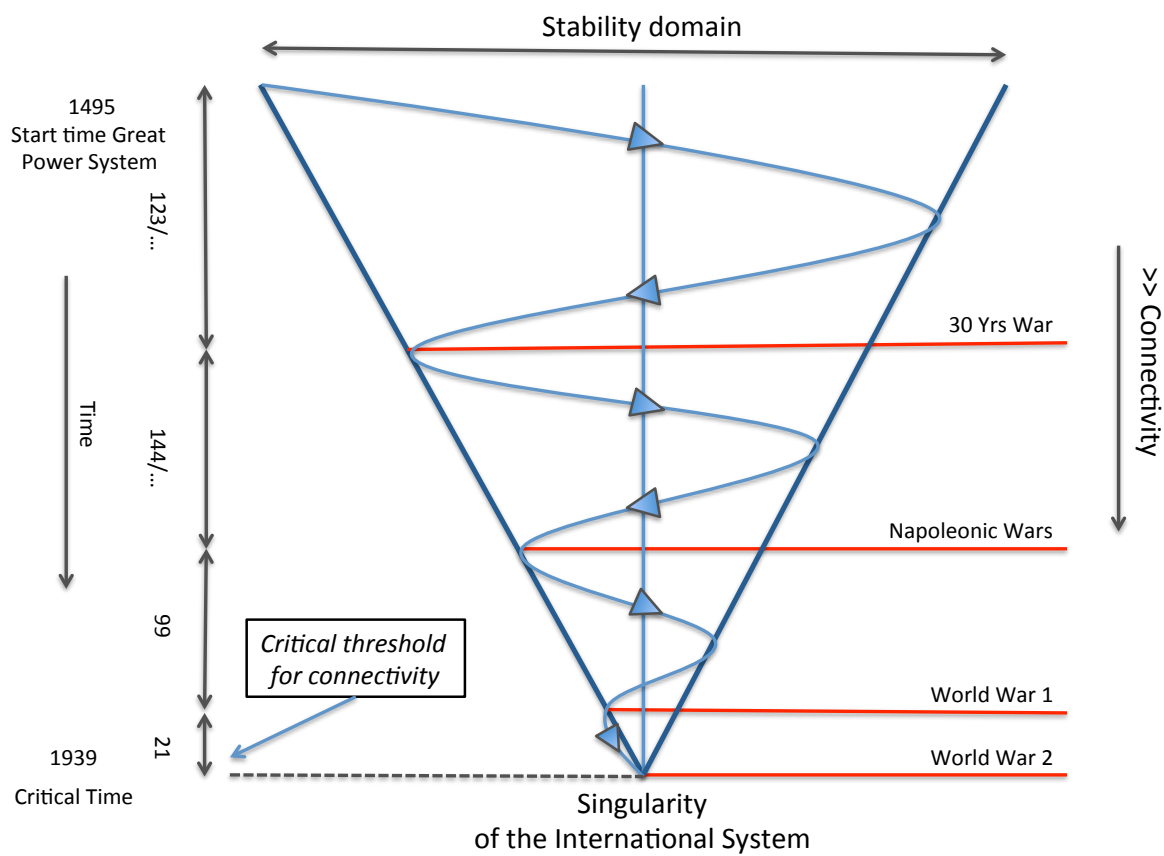

*Figure 5. This image shows and explains the workings of the singularity dynamic, as accompanied by accelerating log-periodic oscillations in the International System, from a different angle. For further explanation, see the text accompanying Figure 3.*

## 6. Connectivity and network effects

*Introduction.* In this section, I discuss the model developed by Watts (55)(56), which provides useful clues for this study. I discuss two cascade (domino-effect) regimes that Watts distinguishes and discuss my interpretation of these regimes.

Going to war is a human decision, despite the fact that decision makers sometimes may feel that they have no choice. Typically, many factors and considerations are simultaneously involved in these situations, and many stakeholders and interests must be taken into account and balanced: the stakes are high. Despite these 'complications,' the problem comes down to a question with a simple structure: 'yes or no we go to war.' These types of decisions are called 'binary decisions with externalities.'

Wars are not initiated with the intention of creating uncontrolled escalation. History – including recent history – shows the impossibility of accurately estimating the outcome and duration of wars. War is unpredictable.

However, not all war decisions lead to escalation, nor do they escalate into systemic wars with far reaching consequences for the International System.

An important question is why, under what conditions, do 'local' decisions result in global – systemic – wars? Why did the shooting of Archduke Franz Ferdinand in Sarajevo on 28 June 1914 trigger a systemic war, whereas various earlier conflicts in the Balkans did not result in such a cascade, such a domino effect? Is there a certain logic to these type of escalations, and can a mechanism be identified that 'facilitates' these escalations?

*Making decisions*. The type of phenomenon – a relatively small trigger causing a massive reaction – is not limited to the phenomenon of war: stock markets occasionally exhibit large fluctuations (crashes) even when no new information becomes available to them. Similarly, innovations and fashion fads exhibit the same patterns: some products become subject to 'hype,' whereas other products with seemingly similar characteristics remain unmoved by 'hype.'

"These phenomena are examples of what economists call information cascades, during which individuals in a population exhibit herd-like behavior because they are making decisions based on the actions of other individuals, rather than relying on their own information about the problem" (55, p5766).

Under certain circumstances, particularly when information is not readily available, other decision makers are watched carefully to decide when and how to act (49, p99). I assume that decision makers deciding about war ('yes or no we go to war') to a high degree base their decisions on the actions of other decision makers (see also Levy, cited in section 3 (23)); frequently, these other decision makers are actually considered the problem. The structure and chronology of the decision-making process in European capitals preceding the outbreak of the First World War show this type of behavior (15).

Watts observes in this respect: ".... just as important as the cascades themselves, is that the very same systems routinely display great stability in the presence of continual small failures and shocks that are at least as large as the shocks that ultimately generate a cascade. Cascades can therefore be regarded as a specific manifestation of the robust yet fragile nature of many complex systems: a system may appear stable for long periods of time and withstand many external shocks (robust), then suddenly and apparently inexplicably exhibit a large cascade (fragile)" (55, p5766).

What adds to this effect – making decisions based on the actions of others – is that time is of the essence in war decisions. If war is considered inevitable, it may be wise to act first. 'Achieving (strategic) surprise' is a principle of war.

In this section, I discuss "*a simple model of global cascades on random networks,*" developed by Watts. Watts explains: "some generic features of cascades can be explained in terms of the connectivity of the network by which influence is transmitted between individuals" (55, p5766). I consider wars in the International System, with varying sizes, the outcome of cascades.

I demonstrate that this model provides useful clues to identify the mechanisms that can explain the war dynamics of the International System.

I assume that the decision-making process concerning war is similar in its structure to the decision-making mechanism described by Watts.

Furthermore, although Watts discusses a 'random' network, the International System does not have a random structure. However, as Watts suggests: "Although random graphs are not considered to be highly realistic models of real-world networks, they are frequently used as first approximations because of their relative tractability, and this tradition is followed here" (55, p5767). I also follow this line of thought.

*The Watts model*. The Watts model is "motivated by considering a population of individuals each of whom must decide between two alternative actions, and whose decisions depend explicitly on the actions of other members of the population. In social and economic systems, decision makers often pay attention to each other either because they have limited information about the problem itself or limited ability to process even the information that is available" (55, p5766). As discussed, I consider the decision-making process concerning 'yes or no we go to war' to belong to this class of 'binary decisions with externalities.'

As Watts shows: "As simplistic as it appears, a binary decision framework is relevant to surprisingly complex problems." The simulations show that in particular three features of the network are essential to the dynamics of the network: local dependencies, fractional thresholds and heterogeneity. Local dependencies refer to the effect that a single node (a state prepared to go to war) will have on a given node, critically depends on the preparedness of the other neighboring nodes to (yes or no) go to war. Local dependencies are closely related to the connectivity of the network.

This dynamic, for example, was at play in the European capitals in the months preceding the outbreak of the First World War: the decisions in Berlin, London, Vienna, and Paris in 1914 – the 'yes or no we go to war' decisions – critically depended on the decisions in the other capitals (15).

*Fractional thresholds* of decision makers define the point at which, when a fraction of a country's neighbors have reached a particular state (yes or no), a node will switch its state from yes to no as well (or vice versa). *Heterogeneity* refers to the fact that different nodes can have different thresholds; these differences have an impact on the characteristics of cascades, such as their size.

Another relevant concept Watts introduces is the concept of 'vulnerable clusters.'

A *vulnerable cluster* is a (sub)network of the nodes of the system in which the nodes are only one step away from a shift in their state (a switch from yes to no, or vice versa). Watts call nodes that are *unstable* in this one-step sense *vulnerable* and those that are not *stable* (55, p5767).

This approach shows that a cascade (a war, for example), to escalate or not to escalate, does "depend less on the number and characteristics" of the decision makers themselves than on the structure of the network of the decision makers, who comprise a vulnerable cluster (one step away to decide to go to war).

Watts assumes that "the required condition for a global cascade is, that the subnetwork of vulnerable nodes must percolate throughout the network as a whole."

Thus, "regardless of how connected the network might be," Watts explains, "only if the largest vulnerable cluster percolates are global cascades possible" (55, p5768). This condition of a network is called a cascade condition, which indicates that regardless of how connected the International System is, systemic wars (a war with the size of the system) are only possible when the 'underlying' cluster of states that are unstable has 'percolated' through the International System.

For a cascade – domino effect – to become global, the network must have a minimum level of connectivity, allowing for vulnerable clusters to percolate through the entire system. This indicates that two regimes can be distinguished based on the level of connectivity of the International System and separated by a phase transition: regimes that can and regimes that cannot 'support' global cascades. It is a characteristic of the network.

*Two regimes.* Simulations show that a cascade window can be identified; outside that window, global cascades are impossible. Two parameters define the window that allows for cascades: (1) the number of connections a node has with other vulnerable nodes (z) and (2) the level of the thresholds that apply to these nodes. In this window, all nodes employ the same decision thresholds (threshold heterogeneity = 0).

Watts explains that "the onset of global cascades can occur in two distinct regimes: in a low connectivity regime (I define as regime 1) and in a high connectivity regime" (I define as regime 2), corresponding to the lower and upper phase transitions, respectively. The nature of the phase transitions at the two boundaries is different: this differentiation has "important consequences for the apparent stability of the systems involved" (55, p5770), including for the International System, as I explain later.

Initially, in regime 1, "the propagation of cascades (decisions to go to war) is constrained principally by the connectivity of the network." The connectivity determines the size of cascades.

I argue that a similar effect – a network effect – can be identified in the war dynamics (cycle level) of non-systemic wars of the International Network; shortly after a systemic war, the connectivity of the International System remains low, the number of issues (vulnerable clusters) remains limited, and wars are relatively small as a consequence. At this stage, the propagation of war is constrained principally by the connectivity of the network. However, as time passes, the connectivity of the network increases and the number of issues (as a consequence of the dynamics on the network, such as differentiated development and growth of Great Powers, as discussed by Gilpin) grows. This dynamic results in larger wars.

However, at the upper boundary, the regime is different (regime 2). Here, "the propagation of cascades is limited not by the connectivity of the network, but by the local stability of the nodes." "Most nodes (issues) in this regime have so many neighbors, that they cannot be toppled by a single neighbor perturbation; hence, most initial shocks immediately encounter stable nodes" (55, p5770). In the context of this study, this connectivity effect – resulting in stability – can be explained as follows: decision makers in international affairs use thresholds to guide their decisions – "if a certain 'border' is crossed, sanctions will follow." If issues are sparsely connected, a single other state, by changing from yes to no (or vice versa) can make a difference. For example, if the 'decision rule' of a particular state is that it changes its readiness to go to war when 3 out of 5 other states connected to a certain issue decide to go to war, one neighboring state can make a difference: 3 out of 5 becomes 4 out of 5, and a decision is made to go to war.

However, if, as Watts explains, a high connectivity regime is in place, then the high connectivity of the network will limit cascades as a consequence of the local stability of the nodes. Even when the same threshold (or decision rule) is in place but the actual ratio is 26/50 (50 states, of which 26 are prepared to go to war), one neighboring state, switching form no to yes, will make less of a difference for the state in question. The state will not change its position and will not go to war.

In this case, in other words, states "have so many 'neighbors' that they cannot be toppled by a single neighbor perturbation; hence, most initial shocks immediately encounter stable vertices. Most cascades therefore die out before spreading very far." A high connectivity, under certain network conditions, has a stabilizing effect. However, this stabilizing effect is only temporary: "A percolating vulnerable cluster still exists and it will be rarely triggered. But if this massive vulnerable cluster is triggered, an extremely large cascade is the result."

So, "as the upper phase transition is approached from below, global cascades become larger, but increasingly rare, until they disappear altogether, implying a discontinuous (i.e., first-order) phase transition in the size of successful cascades. The main consequence of this first-order phase transition is that just inside the boundary of the window, where global cascades occur very rarely, the system will in general be indistinguishable from one that is highly stable, exhibiting only tiny cascades for many initial shocks before generating a massive, global cascade in response to a shock that is a priori indistinguishable from many others."

*Interpretation.* The war dynamics of the International System seem to show a similar dynamic. Following a systemic war, wars are initially relatively small; as time passes, they grow in size and reach a 'turning point' at which they begin decreasing. Typically, the period before the next systemic war is relatively 'stable.' The International System routinely displayed great stability in the presence of continual small shocks, as well. This stability, as the Watts model suggests, is a consequence of the high connectivity of the International System at that stage of its development.

During the decades leading up to the First World War, European countries and their populations held out high expectations for world peace: the system was stable, wars were not frequent. It was argued then – as we argue today – that intense international trade made war impracticable and that war was actually in nobody's interest. However, the Watts model suggests that war – particularly the escalation of war – forms part of the fabric of the International System, and particularly, of its network. It seems that in the years leading up to the First World War, a system-sized vulnerable cluster percolated through the network of the International System and caused a massive cascade that was 'triggered' in the month of June 1914 in Sarajevo.

*These network effects, which are related to the connectivity of the International Network, create a paradox: connectivity – for example, the economic interdependence of states – is sometimes used as a rationale or argument that war is 'not logical' because it is not beneficial for the states involved. To a certain extent, this assumption is 'rational.' However, this study also shows that to fully understand the dynamics on the network of international relations, the characteristics and development of the underlying network must also be taken into account. This study demonstrates that at a certain level of connectivity of the underlying network, conflict becomes more likely because of the lack of regulatory capacity. Thus, connectivity has two paradoxical effects.*

Watts' findings provide important clues for a better understanding of the war dynamics of non-systemic wars (all wars, except for the four wars that I identify as systemic wars). The model and simulations of Watts support the assumption that connectivity drives and shapes the (war) dynamics of the International System, and connectivity is the defining characteristic of the war dynamics of the International System. The typical regularities that Watts describes in the dynamics of cascades seem to be recognizable in the war dynamics of the International System, as well.

In my study, I refer to 'network effects.' These effects concern the impact of the connectivity of the network (1) on the size and frequencies of wars and (2) on how these characteristics evolve during the life span of successive cycles. It seems that successive cycles – the quasi-stable periods between two successive systemic wars – show a typical life cycle that can be explained by the development of the connectivity of these cycles.

**Schematic representation of the Life Span of a Cycle**

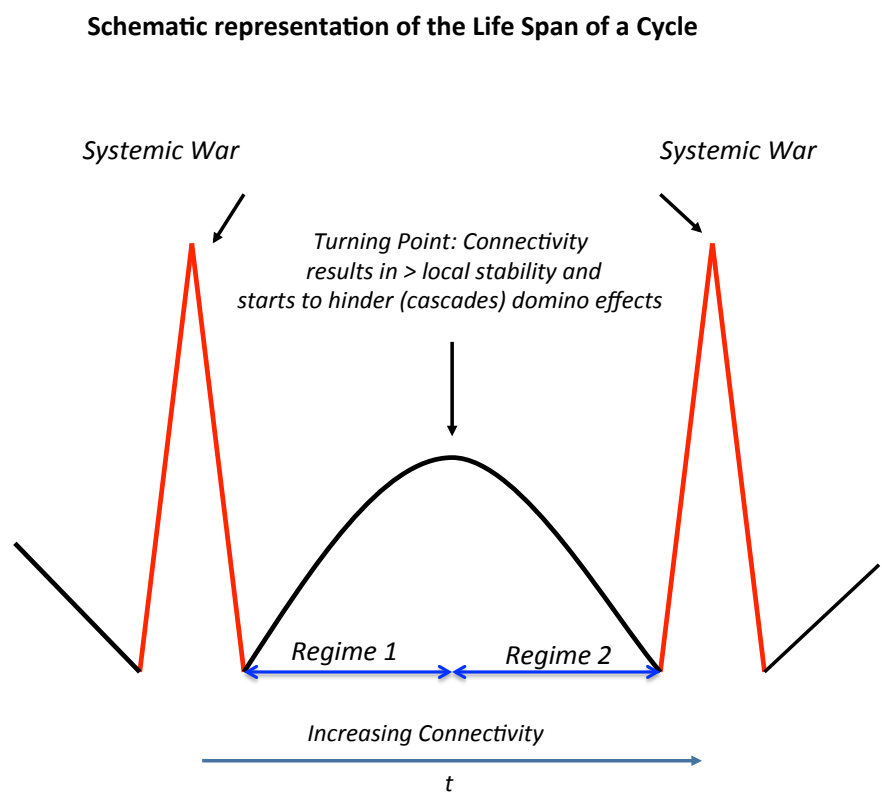

*Figure 6. This figure schematically presents the war dynamics of the International System during the life span of a cycle, i.e., the quasi-stable period between two successive systemic wars. Initially, as the connectivity of the International System increases, the wars grow in size, reach a turning point, and begin declining. The turning point marks the moment in time when the increased connectivity results in increased stability and begins to 'hinder' the war dynamics. The turning point marks the 'switch' from regime 1 to regime 2.*

In section 5, it was suggested that the four systemic wars constitute re-organizations of the regulatory network of the International System: at a certain point, the accelerated growth of the regulatory network cannot be sustained and the network breaks down. In this section, systemic wars are defined as global cascades. The development of the size of these cascades is closely connected to the connectivity growth of the network, as Watts explains.

An important issue to address is how the saturation of the regulatory network – the governance of the International System – 'matches' with the assumptions I make regarding the war dynamics of the International System based on Watts. What is the relationship between the two dynamics? Are these two perspectives 'consistent'?

The global cascade marks – is symptomatic of – the collapse of the regulatory network of the International System. It is the same phenomenon 'approached' from a different analytical point of view. Watts' model helps identify the underlying mechanism for this dynamic from a network perspective.

At the end of the life span of a cycle, the connectivity of the regulatory network of the International System reaches a level that 'creates" the increased local stability of the network. This increased connectivity and local stability – I assume – also reflect the saturation of the regulatory network. Connectivity has reached such a level that effective and timely 'global coordination' (as defined by Mattick et al. (35)) has become impossible. The saturation of the network further fuels the massive global cluster that is 'waiting' to be triggered.

When issues can no longer be assessed on their own merits, decision makers are inclined to focus (even more) on the actions and decisions of (potentially) antagonistic actors: a small incident can now trigger a system-sized domino effect.

The massive global cascade (of Watts (55)) goes hand in hand with – and is symptomatic of – the actual collapse of the regulatory network. The trigger confronts decision makers with a situation that they cannot adequately assess because the regulatory network has already become dysfunctional. The imitation of decisions functions as a positive feedback loop.

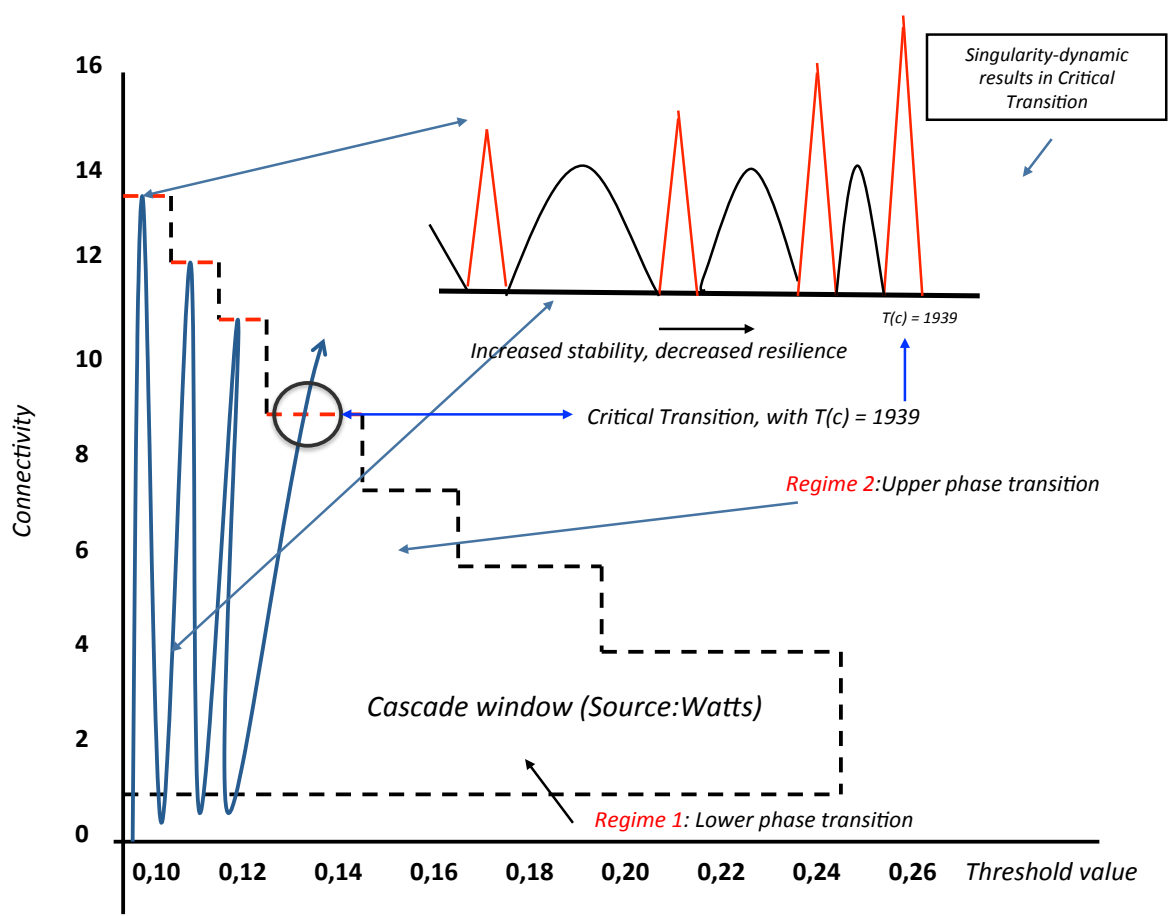

*Figure 7. This figure shows the cascade window developed by Watts. The borders of the window are defined by the connectivity of the network and the thresholds that nodes 'use' in their decision rules to change their 'state.' The two regimes are also shown. I assume that the war dynamics – as it were – oscillate in this window until a critical transition is set in motion that transforms the system fundamentally. The (outcome of the) critical transition anchors the 'new' system above the upper phase transition in which cascades are impossible.*

## 7. War dynamics with chaotic and periodic characteristics

*Introduction.* "Chaos is a phenomenon encountered in science and mathematics wherein a deterministic rule-based system behaves unpredictably. That is, a system, which is governed by fixed, precise rules, nevertheless behaves in a way that is, for all practical purposes, unpredictable in the long run" (20). Chaotic systems are deterministic systems that appear to be random but that in fact follow precise (mathematical) rules (37).

"Mathematically chaotic systems are, in a sense perfectly ordered, despite their apparent randomness." "The study of chaos shows that simple systems can exhibit complex and unpredictable behavior. This realization both suggests limits on our ability to predict certain phenomena and that complex behavior may have a simple explanation" (20).

In this section, I suggest that war dynamics at the cycle level (excluding systemic wars) are (1) partly deterministic (governed by 'simple' rules in which the input determines the output) and show chaotic characteristics and (2) are partly stochastic, which indicates that the war dynamics also have some element of change.

The output of a (deterministic) function in mathematics is used as the input for the next step, which can be thought of as a feedback process in which output is used as input. This process thus results in orbits.

A dynamical system is chaotic if it possesses each of the following properties (20):

- The dynamical rule is deterministic.
- The orbits are aperiodic, i.e., they never repeat.
- The orbits are bounded and thus remain between an upper and lower limit.
- The dynamical system has sensitive dependence for initial conditions. A system that has sensitive dependence for initial conditions has the property that a very small change in the initial condition will lead to a very large change in the orbit in a relatively short period.

Chaos requires a few so-called degrees of freedom – at least two, but not too many. Chaotic dynamics rely on the assumption that only a few major variables interact nonlinearly and create complicated trajectories (49).

It is not possible to scientifically prove that war dynamics are chaotic in this mathematical sense. Insufficient data are available, not all of the data are accurate, and the dynamics are influenced by stochastic events. It is not possible to clearly separate the stochastic from the chaotic components, when the latter component exists in the first place.

In this section, I (1) show the 'construction' of the trajectories of non-systemic wars in the phase state, (2) calculate the Lyapunov exponents of two trajectories, and (3) more closely examine the war dynamics – abnormal war dynamics – during a specific period in time that according to historians, was characterized by intense rivalry between two Great Powers. I argue that this intensity temporarily decreased the degrees of freedom of the International System to two, resulting in more regular – periodic – war dynamics.

*Constructing trajectories in the phase state.* In the phase state, chaotic dynamics show complicated trajectories. These trajectories – and their boundaries – result from the so-called strange attractors of these types of systems. These complicated trajectories are the outcome of a few variables interacting nonlinearly.

The figure below shows the trajectories of the war dynamics of non-systemic wars during the period. In the figure below, the seven orbits that can be identified in the first cycle (1495–1618) are shown. When a phase state is constructed for the war dynamics of the International System, based on the fraction and intensity of successive wars, it is possible to identify circular trajectories (orbits). Some orbits follow left-handed and others right-handed trajectories. The right-handed orbits are projected on the left side (2nd quadrant), and the left-handed orbits are on the right side. The data show that at certain points in time, non-systemic wars do not follow these circular trajectories; apart from the exceptional period, these are only short interruptions.

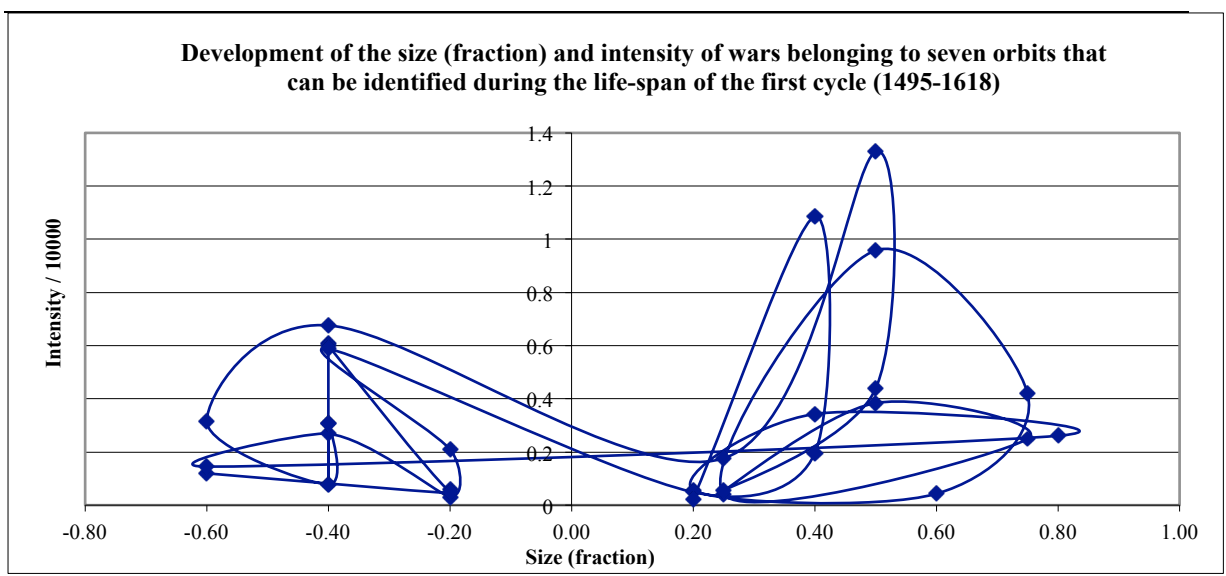

*Figure 8: This figure shows the orbits of successive wars in the phase state belonging to the first seven orbits that can be identified during the life-span of the first cycle (1495–1618), with size (fraction) and intensity as variables. In quadrant I, the orbits with a left-handed direction are shown, and in quadrant II, right-handed orbits. I have constructed this 'attractor' by visually identifying circular orbits in the phase state and determined whether these trajectories follow a left-handed or right-handed trajectory. Next, I have projected the right-handed trajectories in the other (second) quadrant. Further analysis reveals that the wars belonging to a specific orbit do <u>not</u> develop arbitrarily; these groupings develop with a certain regularity and (unknown) logic (see annex). These typical orbits show (visual) similarities with strange attractors, which is typical for chaotic systems. These visual similarities do not prove anything.*

The 'attractor' of the war dynamics during this period of time shows visual similarities with attractors of chaotic systems. However, although this similarity is remarkable, it does not prove anything (52).

*Lyapunov Exponents.* A typical characteristic of chaotic dynamics is their sensitive dependence for initial conditions, which indicates that two almost similar initial conditions will develop differently, which makes accurate longer-term predictions impossible. The Lyapunov exponent is a measure of the rate of spread of two nearby initial conditions. "The Lyapunov exponent is defined as the average rate of trajectory divergence caused by the endogenous component (and not by stochasticity), using for its calculation two trajectories that start near one another and that are – this is an important assumption – affected by an identical sequence of random shocks" (26). A positive exponent is supposedly an indicator of chaos.

To calculate the Lyapunov exponents of the war dynamics of the International System, I have selected 2 pairs of wars (serial 1 and serial 2, respectively), of which the initial conditions in the phase state (size and intensity) are close to one another. I consider the point at which these two wars show almost similar size and intensity 'values' the origin of both trajectories.

A note of caution must be made: the initial conditions for size and intensity are more or less approximate in the phase state, but in 'time' they differ significantly (the time of the respective wars), i.e., pair 9-36 and pair 9-39, are respectively, 74 and 87 years apart. As a consequence, both trajectories (of a single pair) were subject to different levels of noise. However, if this development results in distortions of the calculations, it will most likely be more difficult to identify positive Lyapunov exponents.

Serials 1 and 2 concern war numbers 9 and 39 and war numbers 9 and 36, respectively. Next, I have determined λ for: $|\Delta I(t)| = |\Delta I(0)|e^{\lambda(t)}$.

| Pairs with initial conditions | | | |
|---|---|---|---|
| Serial | Pair | Start value size, both wars | Start value intensity (/1000), war 1 and war 2 |
| 1 | 9–39 | 0,20 | 0,057 and 0,024 |
| 2 | 9–36 | 0,20 | 0,057 and 0,049 |

*Table 4. This table provides information on the selected pairs with almost similar conditions.*

| Serial 1: Lyapunov exponent ≈ 1,02 | | | |
|---|---|---|---|
| t | 9 | 39 | Abs |9-39| | λ |
| 0 | 0,057 | 0,024 | 0,033 | NA |
| 1 | 0,043 | 0,127 | -0,084 | 0,93 |
| 2 | 0,42 | 0,149 | 0,271 | 1,05 |
| 3 | 0,958 | 0,003 | 0,955 | 1,12 |
| 4 | 0,041 | 1,685 | -1,644 | 0,98 |

*Table 5. This table shows the data used for the calculation of the Lyapunov exponent for this pair. The Lyapunov exponent = 1,02*

| Serial 2: Lyapunov exponent ≈ 2,51 | | | |
|---|---|---|---|
| t | 9 | 36 | Abs |6-36| | λ |
| 0 | 0,057 | 0,049 | 0,008 | NA |
| 1 | 0,043 | 0,195 | -0,152 | 2,94 |
| 2 | 0,42 | 1,086 | -0,666 | 2,21 |
| 3 | 0,958 | 0,024 | 0,934 | 2,38 |

*Table 6. This table shows the data used for the calculation of the Lyapunov exponent for this pair. The Lyapunov exponent = 2,51*

These two pairs show a positive Lyapunov exponent, which points to chaotic dynamics. The values of the separate exponents are in relatively close proximity, particularly when different noise levels are taken into account. Not all wars (pairs) in close proximity in the phase state have positive Lyapunov exponents.

These results are encouraging. Apart from visual similarities in the attractor, there are positive Lyapunov exponents for the two pairs of initial conditions. Another clue also sheds light on the characteristics of the non-systemic war dynamics: the abnormal war dynamics during the exceptional period.

*Periodic war dynamics.* When constructing the phase state for the intensity and fraction of all non-systemic wars, it is possible to identify 'abnormal' dynamics during a specific period of time: 1657–1763. These abnormal dynamics during the life span of the second cycle may likely provide additional insights into this question.

According to historians (46), the (war) dynamics during this timeframe were to a high degree dominated by the intense rivalry between Great Britain and France, with each (politically) maneuvering and fighting for a hegemonic position in Europe. During this period of time, wars frequently were either relatively small with a low intensity or large with a high intensity. The rivalry between Great Britain and France resulted in a simple 'zigzag' war dynamic in the phase state. The rivalry between these Great Powers reached its peak in the Seven Years' War and was eventually settled in favor of Great Britain in 1763. After this war, the typical circular trajectories resumed. These changes in dynamics may be defined as 'bifurcations.'

The period from 1763 until the French Revolutionary and Napoleonic wars – the second systemic war – was a relatively quiet period, with only minor conflicts (46). The relatively peaceful and typical end of this life cycle – as is the case with the other cycles – is consistent with the cascade dynamics present shortly before a massive global cascade – a systemic war in this context – is triggered.

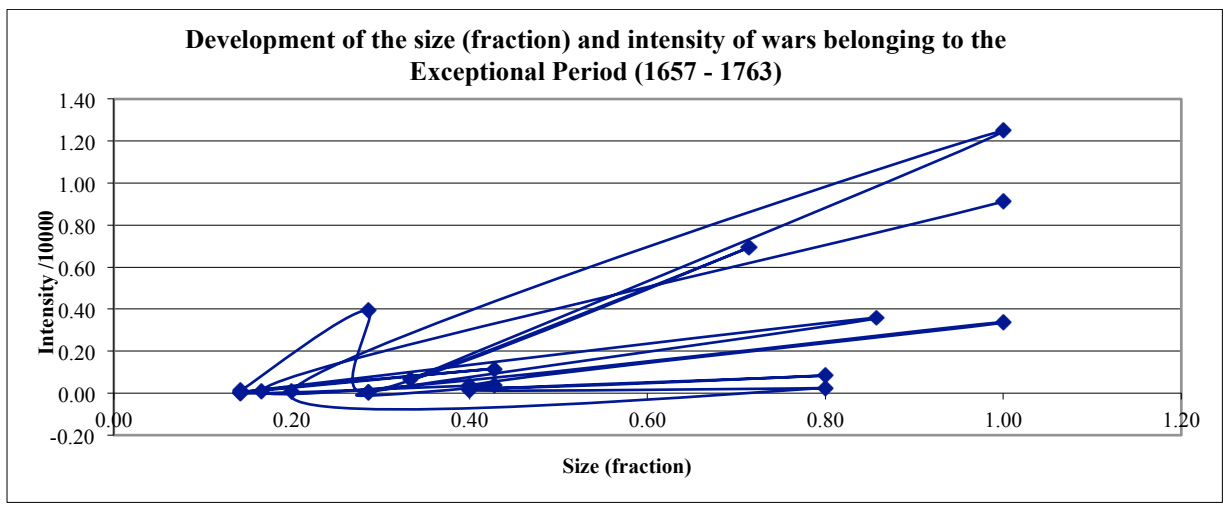

*Figure 9. This figure shows the size and intensity of wars during the exceptional period: the different types of dynamics can be determined visually by comparing the trajectories in Figure 8.*

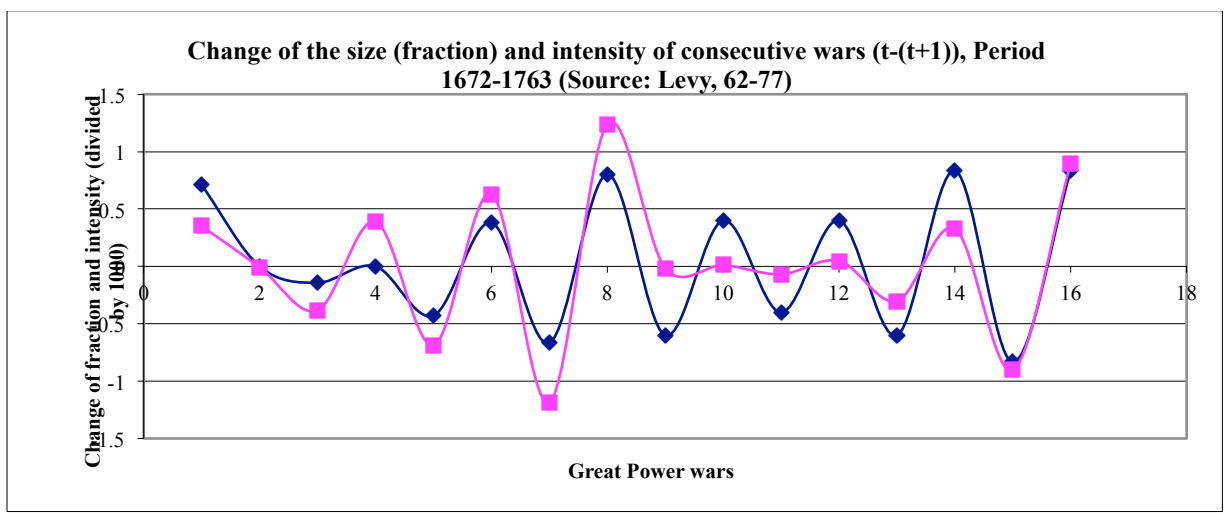

*Figure 10. The 'abnormal' dynamics during the exceptional period show signs of periodicity. The regular development of the change in war size and intensity of wars 62-77 (Levy) shows this feature in particular.*

To obtain a better understanding of this more periodic dynamic, I calculated the autocorrelation of the war data: see the table below. This analysis confirms (1) the fundamentally different dynamics during the exceptional period and (2) that the dynamics during the exceptional period show periodic characteristics.

| Autocorrelation lag-2 | | |
|---|---|---|
| | **Data time series** | **AC lag-2** |
| 1 | All data, until World War 2 (nr. 113 Levy) | 0.098 |
| 2 | Conform 1, wars during exceptional period excluded (nr.'s 58-77 Levy) | - 0.060 |
| 3 | Wars during exceptional period only (58-77 Levy) | 0.486 |

*Table 7. This table shows the outcome of the autocorrelation lag-2 of three datasets. Levy's data contain 119 wars. The following wars are excluded from this overview: 6 wars that occurred between 1945 and 1975; 8 wars that comprised 4 systemic wars; and 20 wars that took place during the exceptional period. The lag-2 autocorrelation may be regarded as moderate.*

I now suggest that during this exceptional period, the degrees of freedom of the International System were temporarily decreased to two due to the intense rivalry between Great Britain and France. Normally, I assume that there are at least three degrees of freedom of the system, providing for a fundamentally different – most likely chaotic – dynamic. Chaotic dynamics require at least three degrees of freedom.

During the exceptional period, all 'international' issues and interactions were to a high degree related to the rivalry between Great Britain and France. 'All' other issues were most likely connected to this rivalry.

This finding does not constitute proof for the existence of chaotic war dynamics, but the arguments add up: the typical circular trajectories, the positive Lyapunov exponents, and the abnormal war dynamics that seem to be related to a decrease in the degrees of freedom of the system together make a powerfully compelling case for the hypothesis that 'normally' the war dynamics showed chaotic characteristics. Thus, the International System is a rule-based deterministic system, with only a few (at least three) variables determining its dynamics. It also indicates that the war dynamics – concerning non-systemic wars – are intrinsically unpredictable (at least) with respect to based on their intensity and size.

Thus, systemic and non-systemic wars not only differ in their function and impact (systemic wars are instrumental in periodic reorganizations of the International System and produce organizational innovations) but also differ in their predictability: systemic wars are highly predictable (in size, duration, and intensity, as discussed below), whereas non-systemic wars are highly unpredictable, depending on the number of degrees of freedom of the system.

*The impact of and on connectivity.* I assume that there was interplay between the abnormal war dynamics and the development of the connectivity of the International System during the life span of the second cycle. My reasoning is as follows.

Based on the Watts model, it is possible to identify for each cycle a 'turning point,' i.e., a point in time during the life span of a cycle when the cascades reached a maximum size because of the change of regime (from regime 1 to 2), i.e., from a regime with low connectivity to a regime with high connectivity. During regime 2 (high connectivity), cascades are 'hampered' by local stability effects.

I have visually determined the turning points of the first three cycles; see the table below.

| Turning points of successive cycles | | |
|---|---|---|
| **Cycle** | **Life span** | **Turning point in war dynamics** |
| 1 | 1495–1618 | 1514 |
| 2 | 1648–1792 | 1774 |
| 3 | 1815–1914 | 1856 |

*Table 8. This table shows the turning points of successive cycles. Cycle 4 is not included because only one non-systemic war occurred during this cycle. The turning points are determined visually. A turning point marks the 'switch' between regime 1 and regime 2 dynamics. Regime 2 dynamics are hampered in their development by the (increasing) local stability of the network.*

This 'analysis' shows that the exceptional period is situated before the turning point of the second cycle, which implies that during the exceptional period, a low-connectivity regime was in place. During such a regime, connectivity determines the sizes of the cascades. At this stage (before the turning point), connectivity does not 'create' local stability and begins (progressively) to hamper the development of cascades. The fact that the exceptional period is situated before the turning point most likely explains the large sizes of wars during this period: the connectivity of the network did not restrict them.

Furthermore, it may be argued that the abnormal dynamics that involve only two degrees of freedom delayed the development of the connectivity of the second cycle. As a consequence, the saturation point of the regulatory network was reached later in time and caused the life span of the second cycle to lengthen. The war dynamics were stuck in the first regime – so to speak – during the exceptional period. After the turning point was passed, the normal – I assume chaotic – war dynamics resumed, creating new connections at a faster pace.

'Theoretically,' the life span of the second cycle is actually too long; the life span should be shorter than the life span of the preceding cycle. This lengthening of the life span may most likely be attributed to the abnormal dynamics.

**Schematic representation of the Life Span of the 2nd Cycle**

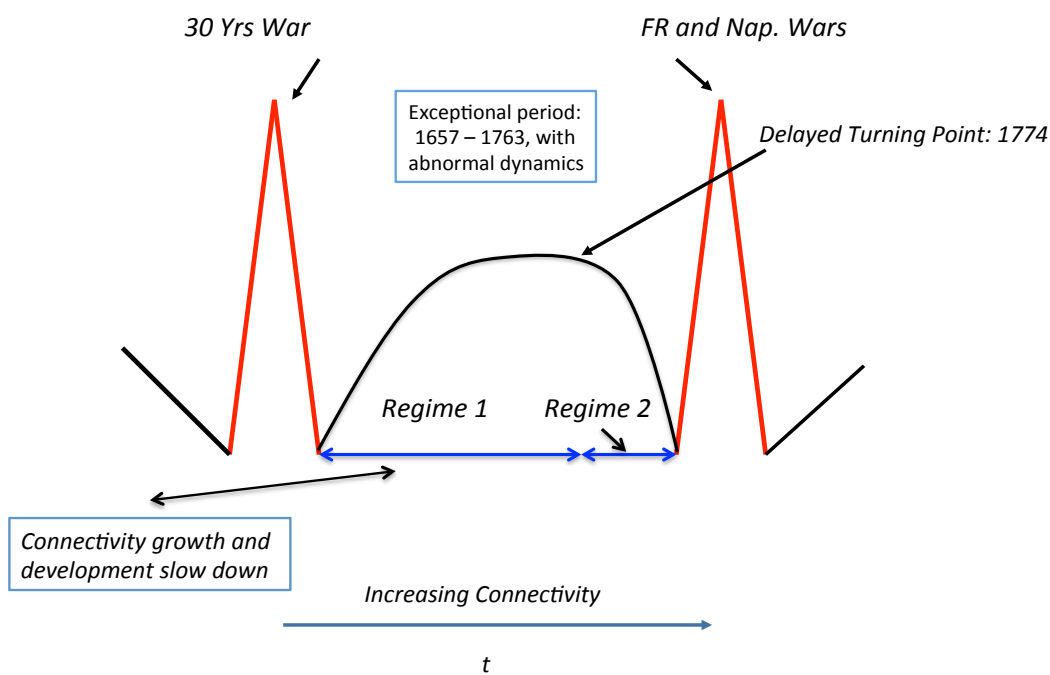

*Figure 11. This figure shows the distorted development of the second cycle. I assume that the International System was temporarily stuck in regime 1 and produced larger wars for some period of time. I assume that a delayed connectivity growth was also an effect of the abnormal war dynamics. It seems plausible that the connectivity growth is delayed with a reduced number of degrees of freedom. This delay caused some 'knock-on' effects: delaying the saturation of the regulatory network of the International System and its eventual collapse. I assume that this distorted development of the second cycle lengthened its life span.*

The relationship between chaotic dynamics and the smooth development of the International System is intriguing. This relationship has been observed before in other systems.

The following observations by Crutchfield (16) regarding complexity and chaos are most likely valid for the dynamics of the International System, as well: "complexity often arises in a middle ground – often at the order-disorder border." "Natural systems that evolve with and learn from interaction with their immediate environment exhibit both structural order and dynamical chaos."

"Chaos, as we now understand it, is the dynamical mechanism by which nature develops constrained and useful randomness. From it follow diversity and the ability to anticipate the uncertain future." "Chaos is the outcome of a combination of factors, but provides the system to select new courses of action." "There is a tendency, whose laws we are beginning to comprehend, for natural systems to balance order and chaos, to move to the interface between predictability and uncertainty. The result is increased structural complexity."

### 8. War categories and characteristics of cycles

*Introduction.* In this section, I define four war categories and examine the characteristics of cycles.

*War categories.* Based on the finite-time singularity and the cycles accompanying this long-term dynamic, it is possible to identify four successive cycles. Each cycle is delimited by a systemic war at the start (except for the first cycle starting approximately 1495) and at the end of its life span.
It is possible to define four types of wars.

| Characteristics of types of wars | | |
|---|---|---|
| | **War category** | **Characteristics** |
| 1 | Systemic wars constituting a critical transition. | System-sized. Highly predictable in time. |
| 2 | Systemic wars, part of log-periodic oscillations, resulting in innovations within the 'parameters' of the current system. | System-sized. Highly predictable in time. Intensity and duration are also highly predictable. |
| 3 | Non-systemic wars with chaotic characteristics. | Inherently unpredictable. |
| 4 | Non-systemic wars with periodic characteristics. | Predictable to a certain degree. |

*Table 9. Four types of war can be distinguished, as is shown in this table. Predictability and effects are distinctive features of these categories.*

*Characteristics of successive cycles.* A closer examination of the statistical characteristics of the wars that occur during the life span of the four successive cycles reveals remarkable patterns in the development of the International System over time:

| Development of the stability and resilience of successive cycles | | | | | |
|---|---|---|---|---|---|
| | | **Stability** | | **Resilience** | |
| Cycle | Period | War frequency | GP Status dynamics | Number of Great Power wars | Life span (years) |
| 1 | 1495 – 1618 | 0.37 | 8 | 45 | 123 |
| 2 | 1648 – 1792 | 0.24 | 5 | 34 | 144 |
| 3 | 1815 – 1914 | 0.18 | 3 | 18 | 99 |
| 4 | 1918 – 1939 | 0.05 | 0 | 1 | 21 |

*Table 10. This table shows the stability and resilience of the four successive cycles; it only addresses non-systemic wars. Systemic wars are not included in this overview because they constitute a fundamentally different category. The number of wars in consecutive international systems and the war frequency, for example, evolve linearly. The war frequency of cycles is calculated by dividing the number of Great Power wars by life span of the cycle. Great Power wars outside the European continent, with only one European participant, are excluded from this overview. Thus, there are seven relevant wars involved (numbers 97, 99, 104, 105, 109, 110, and 111) in the 1856-1941 period. From another perspective, this result shows a different category of wars (wars outside Europe, European Great Powers in war with non-European states). These wars are indicative of the globalization of the International System but obscure the process of social expansion and integration in Europe.*

The *stability* and *resilience* of the International System linearly increase and decrease, respectively. I have defined these two characteristics of the International System as follows (this approach is based on ecosystem research (1)(25)(40): *stability* is the ability of the International System to sustain itself in a condition of rest, that is, in the absence of Great Power wars. The frequency of wars during the life span of each successive cycle is indicative of the stability of these cycles. The frequency of wars during the life span of successive cycles is calculated by dividing the number of wars during the life span of a particular cycle by the duration of its life span. The lifespan of a cycle is the difference between the start year of the systemic war that ended the life span of that cycle and the end year of the preceding systemic war. These calculations demonstrate that war frequencies decrease nearly linearly, which implies – in accordance with this definition of the stability concept – a (nearly) linear increase in the stability of successive cycles.

*Stability1 (t) = War Frequency (t) = 0.465–1.02t, with t is the cycle number, R2 = 0,98*

I define the *status dynamics* of the International System as another indicator of the development of the stability of the International System. The *status dynamics* of the International System is defined as the number of states that acquire or lose their Great Power status. Based on Levy's dataset, it is possible to determine that the status dynamic decreases over time (34, p47). During the first four cycles, eight, five, three, and zero status changes occurred (status changes during systemic wars are excluded), respectively. Two of the three status changes during the third cycle concerned the United States (1898) and Japan (1905). The development of this indicator over time not only emphasizes the increased stability of the European system over time but also signals the increased impact of non-European states on the dynamics of the International System. It may be no coincidence (from a system dynamics perspective) that most status changes occur at the turning point – 'half way' – through the life span of cycles, i.e., when increasing connectivity begins to hinder cascades.

*Stability2 (t) = Status Dynamic (t) = 10.5–2.6t, with t is the cycle number, R2 = 0,99*

I have defined the *resilience* of the International System as the ability of the system to sustain itself within a particular stability domain, a cycle in the context of this study (25).
The number of wars during the life span of a cycle is an indicator of the resilience of the cycle. This analysis shows that the resilience of successive cycles decreased almost linearly over time.

*Resilience (t) = Number of Wars (t) = 61.5–14.8t, with t is the cycle number, R2 =0,99*

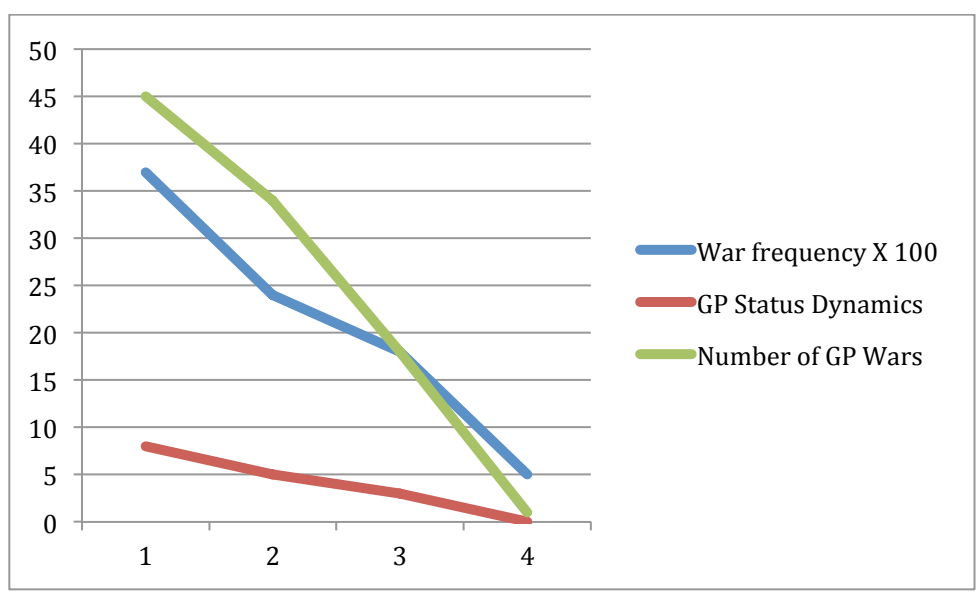

*Figure 12. This figure shows the development of the three variables discussed in this section.*

This study shows that the increase in stability (defined as the number of Great Power wars during the life span of a cycle) is a misleading measure: stability increases, as many historians have noted, but, simultaneously, its resilience decreases. The analysis shows that in the long term, the increased stability of the International System goes hand in hand with the acceleration of the frequency of *systemic* war. Both were typical – and interrelated – developments in the run-up to a destructive singularity, the Second World War, at the critical time (1939) of the singularity. In an anarchistic system, the feeling of safety is an illusion.

*Development of the intensity of systemic wars.* In this section, I take a closer look at the development of the intensity of successive systemic wars. Complex systems sometimes show the combined effect of acceleration in dynamics (shorter cycles) with increased amplitudes. Does the acceleration of oscillations go hand in hand with an increase in the intensities of the systemic wars?

Levy defines *intensity* as "battle deaths per million European population"; thus, it is a ratio.

| | Systemic war | Intensity |
|---|---|---|
| 1 | Thirty Years' War | 23468 |
| 2 | French Revolutionary and Napoleonic Wars | 21928 |
| 3 | First World War | 57616 |
| 4 | Second World War | 93665 |

*Table 11. Systemic wars and their intensities.*

| Development of the intensity and intensity/year of systemic wars | | | | |
|---|---|---|---|---|
| SW | Intensity | Intensity | Intensity/Year | Intensity/Year |
| 1 | 23468 | 23468 | 782 | 782 |
| 2 | 21928 | 21928 | 953 | 953 |
| 3 | 57616 | 57616 | 14404 | 14404 |
| 4 | 93665 | | 18733 | |
| | $9916.24\ e^{\wedge}(0.56\ x)$ | $8664.47\ e^{\wedge}(0.62\ x)$ | $1110.03\ e^{\wedge}(0.72\ x)$ | $5.68\ e^{\wedge}(2.61\ x)$ |
| | $R2 = 0.990$ | $R2 = 0.972$ | $R2 = 0.928$ | $R2 = 0.998$ |

*Table 12. The intensities develop exponentially over time. In particular, the development of the intensity/year shows remarkable regularity.*

The Intensity of successive systemic wars increases exponentially.

| Correlations of systemic war-intensity, and life spans of SW's and cycles | | | |
|---|---|---|---|
| | Intensity SW | Life span of successive systemic wars | Life span of successive cycles |
| 1 | 23468 | 30 | 123 |
| 2 | 21928 | 23 | 144 |
| 3 | 57616 | 4 | 99 |
| 4 | 93665 | 5 | 21 |
| | Correlation: | -0,955* | -0,971 |

*\* This calculation does not include the 4th Systemic War (Second World War) because of its different function.*

*Table 13. This table shows the correlation between the intensity of successive systemic wars and the life span of these wars and between intensities and the life spans of successive cycles.*

The intensity of successive systemic wars is closely (negatively) related to the life span of these wars and to the life spans of successive cycles (independent of the exact scenario). I assume that the connectivity of the International System underlies these correlations.

The life span of successive systemic wars decreases exponentially; I consider this decreasing life span to be a manifestation of the acceleration of the International System that is made possible by the increased connectivity of the system. The increased intensity 'needed' by successive systemic wars most likely indicates the destructive power ('energy') required to 'destroy' (to a certain extent) the old system (i.e., the previous cycle) and subsequently to find a new balance. It seems that more energy is required with the increased stability of the International System. The intensity/year ratio shows a remarkable correlation, in particular.

As discussed above, stability can be defined according to three measures.

The table below shows the correlation between the characteristics discussed in this section.

|   |                     | 1      | 2     | 3     | 4     | 5    | 6    | 7 |
|---|---------------------|--------|-------|-------|-------|------|------|---|
| 1 | Life span of cycles |        |       |       |       |      |      |   |
| 2 | GP status changes   | 0.83   |       |       |       |      |      |   |
| 3 | No. of Non-SW wars  | 0.89   | 0.99  |       |       |      |      |   |
| 4 | War Frequency       | 0.82   | 1.00  | 0.98  |       |      |      |   |
| 5 | Intensity           | - 0.97 | -0.92 | -0.97 | -0.91 |      |      |   |
| 6 | Intensity/Year      | -0.89  | -0.91 | -0.96 | -0.89 | 0.97 |      |   |
| 7 | Life span of SW's   | 0.72   | 0.90  | 0.92  | 0.88  | 0.86 | 0.95 |   |

*Table 14. This table shows a correlation matrix for certain characteristics of the system discussed in this study. I assume that the correlation between the life span of successive cycles and the life span of successive systemic wars is higher (1) when the start date of the first system was established at a somewhat earlier time (36) and (2) when the abnormal dynamics of the second cycle would not have delayed the development of the second cycle.*

This matrix shows that strong (positive or negative) correlations exist between these characteristics. In Figure 12, these statistical relationships are shown.

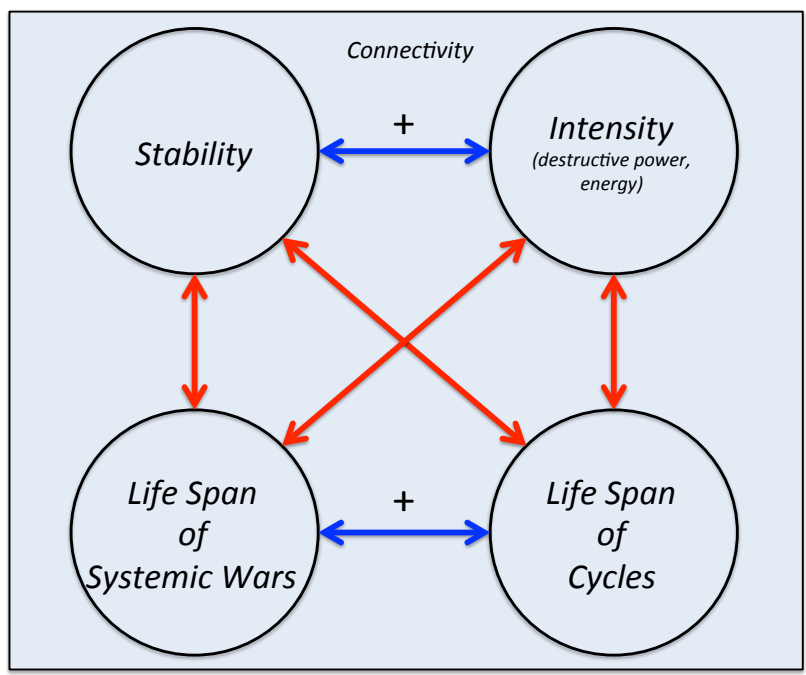

*Figure 13. This figure schematically presents the statistical relationships – correlations – between the developments of certain characteristics of the International System. Red and blue lines show strong negative and positive correlations, respectively. I assume that the causality of these characteristics lies in the connectivity of the International System. This study shows that stability, the intensity of systemic wars, the life span of systemic wars and cycles, and their development over time are closely related to the development of the connectivity of the International System over time.*

## 10. Early-warning signals

This study demonstrates that in the (war) dynamics of the International System and in the development of certain (statistical) characteristics of systemic wars and cycles, significant regularities can be identified. It also becomes clear that the development of the Global System has a clear, self-organized 'direction' toward more cooperation, integration, and differentiation. This direction of development ensures better fulfillment of basic requirements, a greater ability for individuals and social systems to solve problems, and – ultimately – better chances for survival.

These regularities provide clues for early-warning signals for some of the phenomena described in this study: war dynamics, singularities, and the direction of development. However, these regularities only become manifest when the current International System (post-1939-singularity) has the same structural characteristics. I assume that the characteristics that resulted in the singularity dynamic, accompanied by log-periodic oscillations, are fundamentally unchanged: the International System remains anarchistic, and connectivity continues to grow and is most likely growing at an accelerating rate. This acceleration implies that the current system will most likely develop a new – even larger – singularity, and it may be argued that it is already in the process of producing it.

The question is whether early-warning signals may be identified and used to obtain a better understanding of the (war) dynamics of the current International System and to clarify at what stage of development we are in the current cycle. Many clues can be found to identify early-warning signals. I do not identify early-warning signals in this study because this study has a different purpose. I only want to make the case that the development of a set of early-warning signals is not only (technically) possible but also necessary. Because of the immense destructive power that is amassed in the International System, new systemic wars must be avoided, and we must develop the ability to reorganize the International System by means other than systemic war. We now know what the mechanism is that drives us to (systemic) war(s); we now understand that the singularity dynamic is 'pushing' us toward more and 'better' integration. However, this path is destructive and potentially self-destructive. Therefore, we should follow a more sensible path that better ensures our survival and makes more resources available for cooperation (and with more focus) on other global problems, such as climate change and poverty.

Critical transitions in ecosystems and climate change (17) – in addition to early-warning signals that sometimes seem to accompany them – are an important subject for research. This field of research is promising (33)(42)(17).

Some researchers suggest that certain complex systems have generic early-warning signals that make it possible to recognize upcoming critical transitions in advance. Other scientists are more skeptical (27) regarding whether such generic early-warning signals actually exist. The results of this study support this more skeptical view.

The research I refer to demonstrates that when various types of complex systems, such as the climate systems and various ecosystems, approach a tipping point, they show a pattern of critical slowing-down (12). A tipping point sets in motion a critical transition, a dramatic shift in the dynamics of the system. Autocorrelation is considered an indicator – an early-warning signal – for the phenomenon of critical slowing-down.

"Autocorrelation refers to the correlation of a time series with its own past and future values. Also called: 'lagged correlation' or serial correlation, which refers to the correlation between members of a series of numbers in arranged time. Positive autocorrelation might be considered a specific form of persistence, a tendency for a system to remain in the same state from one observation to the next" (43).

"Autocorrelation can be exploited for predictions: an autocorrelated time series is predictable, probabilistically, because future values depend on current and past values."

As discussed above, the dynamics during the exceptional period (1657–1763) show lag-2 autocorrelation, as opposed to the 'normal' war dynamics. This change of dynamic, with moderate lag-2 autocorrelation, does – as I explain – contribute to the lengthening of the life span of this cycle. However, although this dynamic with increased autocorrelation does actually decrease the rate of change, I do not consider this particular instance of autocorrelation to be an indication – or early-warning signal – of an upcoming critical transition (1939): I consider it an anomaly.

As opposed to the system demonstrating the phenomena of critical slowing-down and increased autocorrelation, the International System does not show such a dynamic; instead, its development is accelerating. Critical slowing-down is not a universal property of systems approaching a tipping point.

It is necessary for the identification of a consistent set of early-warning signals to differentiate between dynamics *on* the network of the International System and the underlying dynamics *of* the network. Furthermore, the interplay between these two levels must be considered, in addition to the fundamental difference between systemic and non-systemic wars.

Early-warning signals must be identified and 'devised,' and operating such a set of signals also requires close cooperation in collecting and analyzing the necessary information.

## 11. Recapitulation and implications

In this section, I discuss some of the potential implications of this research. In this study, I have attempted – with the help of insights from complex systems theory and network science – to arrive at a better understanding of the origins and effects of war.

This study demonstrates that a singularity dynamic that is accompanied by accelerating log-periodic oscillations shaped the war dynamics and development of the International System. The driver of these dynamics was a destructive combination of the connectivity growth of the network of the International System, and its anarchistic setting.

Connectivity growth originates in population growth, and the tendency of humans is to organize themselves to fulfill basic requirements, and ultimately, to survive. The result is a positive feedback loop that perpetuates a process of social expansion and integration. This process began thousands of years ago with the formation of small communities to better fulfill the 'basic' requirements of its members. Various factors, particularly the growth of connectivity, resulted in an acceleration of this process.

The singularity dynamic began at the beginning of the Great Power System in approximately 1500. The critical time of the singularity was 1939; at that time, the connectivity of the network of the International System reached a critical threshold. As a result, a critical transition – the Second World War – transformed the European International System from an anarchistic system into a cooperative security community. This systemic war also marked the 'globalization' of the International System: from 1939 onwards, Europe is no longer at the heart of the war dynamics of the International System.

This study demonstrates that the International System is to a high degree a deterministic system: the singularity dynamic suggests not only this result but also the chaotic character of non-systemic wars during the quasi-stable periods between two successive systemic wars.

We do not shape events as much as we believe we do. In fact, we unintentionally create and shape our own context through a multitude of micro-interactions that subsequently result in the emergent macroscopic behavior identified in this study.

A finite-time singularity, accompanied by log-periodic cycles, dominates this emergent macroscopic behavior. Systemic wars not only mark the collapse of the regulatory network of the International System but also provide new solutions for improving cooperation. This singularity dynamic acts as a 'mechanism' that is instrumental in the periodic reorganization of the International System. Each successive systemic war introduces new and progressively intrusive innovative organizing principles.

One commonality shared by the organizing principles introduced by successive systemic wars is that they improved the ability of actors to process information and coordinate their actions. By accepting the sovereignty principle, for example, interactions of states became more predictable and less ambiguous, which improved the capacity of the regulatory network.

This study demonstrates that connectivity growth produces a deadly dynamic in an anarchistic system. This macroscopic behavior not only shapes the war dynamics of the system but also provides the direction and sets the pace of our interactions. The outcome of this critical transition – a fundamental shift in the direction of cooperation – was in a sense already 'in the making' through the long-term dynamic of the singularity.

The log-periodic oscillations – all the systemic wars – resulted in new, but temporary, arrangements that 'structurally' improved the ability to cooperate in the International System. These cycles can be understood as ever-faster cycles of organizational innovation, and from a systems perspective, as a reset of parameters. However, given the anarchistic context and growing connectivity, these solutions only worked temporarily. Ultimately, when the critical time (1939) was reached, i.e., the connectivity of the network reached a critical threshold, a critical transition – a fundamental reorganization – became inevitable.

The new innovation – cooperation – was from the start embedded in the very existence of the singularity (S). Organizational innovation to improve cooperation actually fueled the singularity. The singularity dynamic was also shown to be robust: at all stages of its development, it was cooperation that was 'selected' as the preferred mode of interaction between states. Obviously, cooperation has much to offer.

The robustness of the singularity is also demonstrated by its consistent development in time. It seems that only the non-chaotic dynamics during the exceptional period could cause a delay in the timing of the next systemic war.

The International System is a subsystem of what may be called the more comprehensive Global System. The problem with the International system is that it is anarchistic, and gives rise to conflict interactions. As the finite-time singularity and log-periodic oscillations show, cooperation is thus far stronger than conflict: each systemic war in fact constitutes a reorganization of the International System, and by means of these systemic wars, a more comprehensive organization principle is introduced each time. Ultimately, by means of a critical transition, Europe was 'transformed' from an anarchistic system into a security community, in which Europeans came to understand that common social problems must and can be resolved by processes of peaceful change. These cooperative principles are partially embedded in Europe's structures at this stage.

The long-term singularity dynamic shaped the development of the International System in various fundamental ways that still must be explored.

First, this study makes us more cognizant of the origins and effects of wars. Furthermore, it provides us with clues and a clear direction for the design of an effective International System – an International System that does not have the shortcomings of the current system. For example, this new system must be 'organized' – by introducing the right design principles, saturation of the regulatory network can be avoided, which is, above all, a political issue.

In addition, this study informs us that it is necessary to re-evaluate certain historical events and historical processes (51). For example, this study demonstrates that wars – systemic and non-systemic – have an endogenous (internal) origin and that exogenous, or external, shocks only serve as a triggering mechanism.

Furthermore, it is crucial to make a distinction between the dynamics *on* the network – on which historians focus – and the dynamics *of* the underlying network. It is also important to acquire a better understanding of the interplay between these two dynamics. The dynamics on the network of the International System are typically unique, whereas the dynamics of the underlying network exhibit remarkable regularities.

The interplay between both levels is evident, for example, in the case of the abnormal war dynamics during the exceptional period as a consequence of the intense rivalry between Great Britain and France. This rivalry temporarily reduced the degrees of freedom to two, hindering a more chaotic dynamic and consequently resulting in a series of large-scale, high-intensity wars. The dynamics had an interesting impact on the connectivity of the underlying network, preventing a network effect from containing these conflicts.

However, it is foremost the dynamics *of* the underlying network that shape the dynamics *on* the network: the historical events 'playing out' on the network are unique (4) but the underlying dynamic is not. The Second World War – according to this study – would have occurred 'anyway,' but is may have occurred with other actors as its main players, for example. We are not in control of the system as much as we think we are: the underlying dynamics shape social forces, and events.

It is likely that the current International System functions as its predecessor did: the system remains anarchistic, and connectivity is still growing. The key elements seem to be in place to start a new finite-time singularity dynamic that most likely had already been set in motion. The consequences of another systemic war, or series of such wars, would be disastrous: the proliferation of nuclear weapons, widespread terrorism, and the global reach of terrorists – to name but a few challenges – make systemic wars even more disastrous. Destruction and fragmentation of the Global System – or its support system (the climate) – cannot be excluded as realistic scenarios if a singularity dynamic cannot be stopped. It is much wiser if we arrive at improved cooperation by means other than systemic war. We should use our combined energy to tackle large-scale problems (such as climate change) and not wait until the damage is done and it is too late.

The outcome of the Second World War for Europe is (at least in retrospect) a happy ending that was facilitated by the intense rivalry between East and West (21)(41), with the Soviet Union and the United States as the dominating protagonists, respectively. This rivalry – in combination with common sense on both sides – made (nuclear) wars 'impossible.' This stalemate provided Europe with a window of opportunity to consolidate and embed various forms of cooperation in its (governance) structure. Recent history shows this process to be laborious and exhaustive. Since the early 1990s, the (war) dynamics – with chaotic characteristics – have gathered pace. The question is, if – and how – much resilience this new structure and new European regulatory network has to absorb wars and other shocks.

As discussed above, the ways in which we think about international politics, international relations, and war need serious reconsideration to avoid a 'war trap' that is embedded in the dynamics of the system itself.

This study shows that we must make fundamental changes in our ways of thinking and the decisions we make. History shows that otherwise, we will remain trapped in a destructive dynamic. Now, with new clues available regarding the origins and effects of war, we do not have a credible excuse for not acting and finding better ways to achieve cooperation.

**Annex: Supporting Information**

This annex provides supporting information. The following information is included:

- Levy's dataset, supplemented with 'size data' (size defined as fraction).
- Development of the size of wars over time.
- The data used to identify the chaotic dynamics of non-systemic wars.

**1. Levy's dataset.**

The table below shows Levy's dataset, including fraction (measure for size).

Levy has defined (p81, p92) the units of measurement as follows: *Duration*: years; *Extent*: number of Powers; *Magnitude*: nation-years; *Severity*: the number of battle-connected deaths of military personnel; *Intensity*: battle deaths per million European population; and *Concentration*: battle fatalities per nation year.
*Fraction* is the unit of measurement I have introduced for the size of the war. It is measured relative to the size of the Great Power System at that moment in time. This measure is calculated by dividing the number of Great Powers involved in a war by the total number of Great Powers that exist at that moment in time.

The wars that are marked red constitute the systemic wars. Numbers 46-49 are the Thirty Years' War, 84-85 the French Revolutionary and Napoleonic Wars, 107 the First World War, and 113 the Second World War. The wars numbered 58–77 constitute the exceptional period (1657–1763) with abnormal war dynamics.

| Nr. Levy | Start | End | Duration | Number GP | Extent | Fraction | Magnitude | Concentration | Intensity | Severity |
|---|---|---|---|---|---|---|---|---|---|---|
| 1 | 1495 | 1497 | 2,0 | 5 | 3 | 0,60 | 1,20 | 1333 | 119 | 8000 |
| 2 | 1497 | 1498 | 1,0 | 5 | 1 | 0,20 | 0,20 | 3000 | 45 | 3000 |
| 3 | 1499 | 1503 | 4,0 | 5 | 1 | 0,20 | 0,80 | 1000 | 60 | 4000 |
| 4 | 1499 | 1500 | 1,0 | 5 | 1 | 0,20 | 0,20 | 2000 | 29 | 2000 |
| 5 | 1501 | 1504 | 3,0 | 5 | 2 | 0,40 | 1,20 | 3600 | 269 | 18000 |
| 6 | 1508 | 1509 | 1,0 | 5 | 3 | 0,60 | 0,60 | 3333 | 145 | 10000 |
| 7 | 1511 | 1514 | 3,0 | 5 | 4 | 0,80 | 2,40 | 1500 | 261 | 18000 |
| 8 | 1512 | 1519 | 7,0 | 5 | 2 | 0,40 | 2,80 | 1714 | 343 | 24000 |
| 9 | 1513 | 1515 | 2,0 | 5 | 1 | 0,20 | 0,40 | 2000 | 57 | 4000 |
| 10 | 1515 | 1515 | 0,5 | 5 | 3 | 0,60 | 0,30 | 2000 | 43 | 3000 |
| 11 | 1521 | 1526 | 5,0 | 4 | 3 | 0,75 | 3,75 | 2000 | 420 | 30000 |
| 12 | 1521 | 1531 | 10,0 | 4 | 2 | 0,50 | 5,00 | 3400 | 958 | 68000 |
| 13 | 1522 | 1523 | 1,0 | 4 | 1 | 0,25 | 0,25 | 3000 | 41 | 3000 |
| 14 | 1526 | 1529 | 3,0 | 4 | 3 | 0,75 | 2,25 | 2250 | 249 | 18000 |
| 15 | 1532 | 1535 | 3,0 | 4 | 2 | 0,50 | 1,50 | 4667 | 384 | 28000 |
| 16 | 1532 | 1534 | 2,0 | 4 | 1 | 0,25 | 0,50 | 2000 | 55 | 4000 |
| 17 | 1536 | 1538 | 2,0 | 4 | 2 | 0,50 | 1,00 | 8000 | 438 | 32000 |
| 18 | 1537 | 1547 | 10,0 | 4 | 2 | 0,50 | 5,00 | 4850 | 1329 | 97000 |
| 19 | 1542 | 1550 | 8,0 | 4 | 1 | 0,25 | 2,00 | 1625 | 176 | 13000 |
| 20 | 1542 | 1544 | 2,0 | 4 | 2 | 0,50 | 1,00 | 11750 | 629 | 47000 |
| 21 | 1544 | 1546 | 2,0 | 4 | 2 | 0,50 | 1,00 | 2000 | 107 | 8000 |
| 22 | 1549 | 1550 | 1,0 | 4 | 2 | 0,50 | 0,50 | 3000 | 79 | 6000 |
| 23 | 1551 | 1556 | 5,0 | 4 | 2 | 0,50 | 2,50 | 4400 | 578 | 44000 |
| 24 | 1552 | 1556 | 4,0 | 4 | 2 | 0,50 | 2,00 | 6375 | 668 | 51000 |
| 25 | 1556 | 1562 | 6,0 | 5 | 2 | 0,40 | 2,40 | 4333 | 676 | 52000 |
| 26 | 1556 | 1559 | 3,0 | 5 | 3 | 0,60 | 1,80 | 3000 | 316 | 24000 |
| 27 | 1559 | 1560 | 1,0 | 5 | 2 | 0,40 | 0,40 | 4000 | 78 | 6000 |
| 28 | 1559 | 1564 | 5,0 | 5 | 2 | 0,40 | 2,00 | 2400 | 310 | 24000 |
| 29 | 1562 | 1564 | 2,0 | 5 | 2 | 0,40 | 0,80 | 1500 | 77 | 6000 |
| 30 | 1565 | 1568 | 3,0 | 5 | 2 | 0,40 | 1,20 | 4000 | 306 | 24000 |
| 31 | 1569 | 1580 | 11,0 | 5 | 2 | 0,40 | 4,40 | 2182 | 608 | 48000 |
| 32 | 1576 | 1583 | 7,0 | 5 | 2 | 0,40 | 2,80 | 3429 | 600 | 48000 |
| 33 | 1579 | 1581 | 2,0 | 5 | 1 | 0,20 | 0,40 | 2000 | 50 | 4000 |
| 34 | 1583 | 1590 | 7,0 | 5 | 1 | 0,20 | 1,40 | 2429 | 210 | 17000 |
| 35 | 1585 | 1604 | 19,0 | 5 | 2 | 0,40 | 7,60 | 1263 | 588 | 48000 |
| 36 | 1587 | 1588 | 1,0 | 5 | 1 | 0,20 | 0,20 | 4000 | 49 | 4000 |
| 37 | 1589 | 1598 | 9,0 | 5 | 2 | 0,40 | 3,60 | 889 | 195 | 16000 |
| 38 | 1593 | 1606 | 13,0 | 5 | 2 | 0,40 | 5,20 | 3462 | 1086 | 90000 |
| 39 | 1600 | 1601 | 1,0 | 5 | 1 | 0,20 | 0,20 | 2000 | 24 | 2000 |
| 40 | 1610 | 1614 | 4,0 | 6 | 2 | 0,33 | 1,33 | 1875 | 175 | 15000 |
| 41 | 1615 | 1618 | 3,0 | 6 | 1 | 0,17 | 0,50 | 2000 | 70 | 6000 |
| 42 | 1615 | 1617 | 2,0 | 6 | 1 | 0,17 | 0,33 | 1000 | 23 | 2000 |
| 43 | 1617 | 1621 | 4,0 | 7 | 1 | 0,14 | 0,57 | 1250 | 58 | 5000 |
| 44 | 1618 | 1619 | 1,0 | 7 | 2 | 0,29 | 0,29 | 3000 | 69 | 6000 |
| 45 | 1618 | 1621 | 3,0 | 7 | 1 | 0,14 | 0,43 | 5000 | 173 | 15000 |
| 46 | 1618 | 1625 | 7,0 | 7 | 4 | 0,57 | 4,00 | 20267 | 3535 | 304000 |
| 47 | 1625 | 1630 | 5,0 | 7 | 6 | 0,86 | 4,29 | 11615 | 3432 | 302000 |
| 48 | 1630 | 1635 | 5,0 | 7 | 4 | 0,57 | 2,86 | 15700 | 3568 | 214000 |
| 49 | 1635 | 1648 | 13,0 | 7 | 5 | 0,71 | 9,29 | 17708 | 12933 | 1151000 |
| 50 | 1642 | 1668 | 26,0 | 7 | 1 | 0,14 | 3,71 | 3077 | 882 | 80000 |
| 51 | 1645 | 1664 | 19,0 | 7 | 1 | 0,14 | 2,71 | 3790 | 791 | 72000 |
| 52 | 1648 | 1659 | 11,0 | 7 | 2 | 0,29 | 3,14 | 4909 | 1187 | 108000 |
| 53 | 1650 | 1651 | 1,0 | 7 | 1 | 0,14 | 0,14 | 2000 | 22 | 2000 |
| 54 | 1652 | 1655 | 3,0 | 7 | 2 | 0,29 | 0,86 | 4333 | 282 | 26000 |
| 55 | 1654 | 1660 | 6,0 | 7 | 3 | 0,43 | 2,57 | 1833 | 238 | 22000 |
| 56 | 1656 | 1659 | 3,0 | 7 | 2 | 0,29 | 0,86 | 2500 | 161 | 15000 |
| 57 | 1657 | 1661 | 4,0 | 7 | 1 | 0,14 | 0,57 | 1000 | 43 | 4000 |
| 58 | 1657 | 1664 | 7,0 | 7 | 3 | 0,43 | 3,00 | 8385 | 1170 | 109000 |
| 59 | 1665 | 1666 | 1,0 | 7 | 1 | 0,14 | 0,14 | 1000 | 11 | 2000 |
| 60 | 1665 | 1667 | 2,0 | 7 | 3 | 0,43 | 0,86 | 6167 | 392 | 37000 |

| Nr. Levy | Start | End | Duration | Number GP | Extent | Fraction | Magnitude | Concentration | Intensity | Severity |
|---|---|---|---|---|---|---|---|---|---|---|
| 61 | 1667 | 1668 | 1,0 | 7 | 2 | 0,29 | 0,29 | 2000 | 42 | 4000 |
| 62 | 1672 | 1678 | 6,0 | 7 | 6 | 0,86 | 5,14 | 10364 | 3580 | 342000 |
| 63 | 1672 | 1676 | 4,0 | 7 | 1 | 0,14 | 0,57 | 1250 | 52 | 5000 |
| 64 | 1677 | 1681 | 4,0 | 7 | 1 | 0,14 | 0,57 | 3000 | 125 | 12000 |
| 65 | 1682 | 1699 | 17,0 | 7 | 2 | 0,29 | 4,86 | 11294 | 3954 | 384000 |
| 66 | 1683 | 1684 | 1,0 | 7 | 2 | 0,29 | 0,29 | 2500 | 51 | 5000 |
| 67 | 1688 | 1697 | 9,0 | 7 | 5 | 0,71 | 6,43 | 15111 | 6939 | 680000 |
| 68 | 1700 | 1721 | 21,0 | 6 | 2 | 0,33 | 7,00 | 2370 | 640 | 64000 |
| 69 | 1701 | 1713 | 12,0 | 6 | 5 | 0,83 | 10,00 | 20850 | 12490 | 1251000 |
| 70 | 1716 | 1718 | 2,0 | 5 | 1 | 0,20 | 0,40 | 5000 | 98 | 10000 |
| 71 | 1718 | 1720 | 2,0 | 5 | 4 | 0,80 | 1,60 | 3125 | 245 | 25000 |
| 72 | 1726 | 1729 | 3,0 | 5 | 2 | 0,40 | 1,20 | 2500 | 144 | 15000 |
| 73 | 1733 | 1738 | 5,0 | 5 | 4 | 0,80 | 4,00 | 4400 | 836 | 88000 |
| 74 | 1736 | 1739 | 3,0 | 5 | 2 | 0,40 | 1,20 | 6333 | 359 | 38000 |
| 75 | 1739 | 1748 | 9,0 | 6 | 6 | 1,00 | 9,00 | 8159 | 3379 | 359000 |
| 76 | 1741 | 1743 | 2,0 | 6 | 1 | 0,17 | 0,33 | 5000 | 94 | 10000 |
| 77 | 1755 | 1763 | 8,0 | 6 | 6 | 1,00 | 8,00 | 26105 | 9118 | 992000 |
| 78 | 1768 | 1774 | 6,0 | 6 | 1 | 0,17 | 1,00 | 2333 | 127 | 14000 |
| 79 | 1768 | 1772 | 4,0 | 6 | 1 | 0,17 | 0,67 | 3500 | 149 | 14000 |
| 80 | 1778 | 1779 | 1,0 | 6 | 2 | 0,33 | 0,33 | 150 | 3 | 300 |
| 81 | 1778 | 1784 | 6,0 | 6 | 3 | 0,50 | 3,00 | 2267 | 304 | 34000 |
| 82 | 1787 | 1792 | 5,0 | 6 | 2 | 0,33 | 1,67 | 192000 | 1685 | 192000 |
| 83 | 1788 | 1790 | 2,0 | 6 | 1 | 0,17 | 0,33 | 1500 | 26 | 3000 |
| 84 | 1792 | 1802 | 10,0 | 6 | 6 | 1,00 | 10,00 | 13000 | 5816 | 663000 |
| 85 | 1803 | 1815 | 12,0 | 6 | 6 | 1,00 | 12,00 | 32224 | 16112 | 1869000 |
| 86 | 1806 | 1812 | 6,0 | 6 | 2 | 0,33 | 2,00 | 6429 | 388 | 45000 |
| 87 | 1808 | 1809 | 1,5 | 5 | 1 | 0,20 | 0,30 | 4000 | 51 | 6000 |
| 88 | 1812 | 1814 | 2,5 | 5 | 1 | 0,20 | 0,50 | 1600 | 34 | 4000 |
| 89 | 1815 | 1815 | 0,5 | 5 | 1 | 0,20 | 0,10 | 10000 | 17 | 2000 |
| 90 | 1823 | 1823 | 0,9 | 5 | 1 | 0,20 | 0,18 | 667 | 3 | 400 |
| 91 | 1827 | 1827 | 0,1 | 5 | 3 | 0,60 | 0,06 | 1800 | 2 | 180 |
| 92 | 1828 | 1829 | 1,0 | 5 | 1 | 0,20 | 0,20 | 35714 | 415 | 50000 |
| 93 | 1848 | 1849 | 1,0 | 5 | 1 | 0,20 | 0,20 | 5600 | 45 | 5600 |
| 94 | 1849 | 1849 | 1,2 | 5 | 1 | 0,20 | 0,24 | 2083 | 20 | 2500 |
| 95 | 1849 | 1849 | 0,2 | 5 | 2 | 0,40 | 0,08 | 1500 | 4 | 600 |
| 96 | 1853 | 1856 | 2,4 | 5 | 3 | 0,60 | 1,44 | 35000 | 1743 | 217000 |
| 97 | 1856 | 1857 | 0,4 | 5 | 1 | 0,20 | 0,08 | 1250 | 4 | 500 |
| 98 | 1859 | 1859 | 0,2 | 5 | 2 | 0,40 | 0,08 | 50000 | 159 | 20000 |
| 99 | 1862 | 1867 | 4,8 | 6 | 1 | 0,17 | 0,80 | 1667 | 64 | 8000 |
| 100 | 1864 | 1864 | 0,5 | 6 | 2 | 0,33 | 0,17 | 1500 | 12 | 1500 |
| 101 | 1866 | 1866 | 0,1 | 6 | 3 | 0,50 | 0,05 | 113333 | 270 | 34000 |
| 102 | 1870 | 1871 | 0,6 | 6 | 2 | 0,33 | 0,20 | 150000 | 1415 | 180000 |
| 103 | 1877 | 1878 | 0,7 | 6 | 1 | 0,17 | 0,12 | 171429 | 935 | 120000 |
| 104 | 1884 | 1885 | 1,0 | 6 | 1 | 0,17 | 0,17 | 2100 | 16 | 2100 |
| 105 | 1904 | 1905 | 1,6 | 7 | 1 | 0,14 | 0,23 | 28125 | 339 | 45000 |
| 106 | 1911 | 1912 | 1,1 | 8 | 1 | 0,13 | 0,14 | 5454 | 45 | 6000 |
| 107 | 1914 | 1918 | 4,3 | 8 | 8 | 1,00 | 4,30 | 258672 | 57616 | 7734300 |
| 108 | 1918 | 1921 | 3,0 | 7 | 5 | 0,71 | 2,14 | 385 | 37 | 5000 |
| 109 | 1931 | 1933 | 1,4 | 7 | 1 | 0,14 | 0,20 | 7143 | 73 | 10000 |
| 110 | 1935 | 1936 | 0,6 | 7 | 1 | 0,14 | 0,09 | 6667 | 29 | 4000 |
| 111 | 1937 | 1941 | 4,4 | 7 | 1 | 0,14 | 0,63 | 56819 | 1813 | 250000 |
| 112 | 1939 | 1939 | 0,4 | 7 | 2 | 0,29 | 0,11 | 22857 | 116 | 16000 |
| 113 | 1939 | 1945 | 6,0 | 7 | 7 | 1,00 | 6,00 | 462439 | 93665 | 12948300 |
| 114 | 1939 | 1940 | 0,3 | 7 | 1 | 0,14 | 0,04 | 166667 | 362 | 50000 |
| 115 | 1950 | 1953 | 3,1 | 5 | 4 | 0,80 | 2,48 | 84510 | 6821 | 954960 |
| 116 | 1956 | 1956 | 0,1 | 6 | 1 | 0,17 | 0,02 | 70000 | 50 | 7000 |
| 117 | 1956 | 1956 | 0,1 | 6 | 2 | 0,33 | 0,03 | 300 | 0 | 30 |
| 118 | 1962 | 1962 | 0,1 | 6 | 1 | 0,17 | 0,02 | 5000 | 1 | 500 |
| 119 | 1965 | 1973 | 8,0 | 6 | 1 | 0,17 | 1,33 | 7000 | 90 | 56000 |

| | | | | | | |
|---|---|---|---|---|---|---|
| 1 | War of the League of Venice* | 41 | Austro-Venetian War | 81 | War of the American Revolution* |
| 2 | Polish-Turkish War | 42 | Spanish-Savoian War | 82 | Ottoman War |
| 3 | Venitian-Turkish War | 43 | Spanish-Venetian War | 83 | Russo-Swedish War |
| 4 | First Milanese War | 44 | Spanish-Turkish War* | 84 | French Revolutionary Wars* |
| 5 | Neapolitan War* | 45 | Polish-Turkish War | 85 | Napoleonic Wars* |
| 6 | War of the Cambrian League | 46 | Thirty Year's War - Bohemian* | 86 | Russo-Turkish War |
| 7 | War of the Holy League* | 47 | Thirty Year's War - Danish* | 87 | Russo-Swedish War |
| 8 | Austro-Turkish War* | 48 | Thirty Year's War - Swedish* | 88 | War of 1812 |
| 9 | Scottish War | 49 | Thirty Year's War - Swedish-French* | 89 | Neapolitan War |
| 10 | Second Milanese War* | 50 | Spanish-Portuguese War | 90 | Franco-Spanish War |
| 11 | First War of Charles V* | 51 | Turkish-Venetian War | 91 | Navarino Bay |
| 12 | Ottoman War* | 52 | Franco-Spanish War* | 92 | Russo-Turkish War |
| 13 | Scottish War | 53 | Scottish War | 93 | Austro-Sardinian War |
| 14 | Second War of Charles V* | 54 | Anglo-Dutch Naval War* | 94 | First Schleswig-Holstein War |
| 15 | Ottoman War* | 55 | Great Northern War* | 95 | Roman Republic War |
| 16 | Scottish War | 56 | English-Sopanish War* | 96 | Crimean War* |
| 17 | Third War of Charles V* | 57 | Dutch-Portuguese War | 97 | Anglo-Perian War |
| 18 | Ottoman War* | 58 | Ottoman War* | 98 | War of Italian Uniffication* |
| 19 | Scottish War | 59 | Sweden-Bremen War | 99 | Franco-Mexican War |
| 20 | Fourth War of Charles V* | 60 | Anglo-Dutch Naval War* | 100 | Second Schleswig-Holstein War |
| 21 | Siege of Boulogne* | 61 | Devolutionary War* | 101 | Austro-Prussian War* |
| 22 | Arundel's Rebellion* | 62 | Dutch War of Louis XIV* | 102 | Franco-Prusssian War* |
| 23 | Ottoman War* | 63 | Turkish-Polish War | 103 | Russo-Turkish War |
| 24 | Fifth War of Charles V* | 64 | Russo-Turkish War | 104 | Sino-French War |
| 25 | Austro-Turkish War* | 65 | Ottoman War* | 105 | Russo-Japanese War |
| 26 | Franco_Spanish War* | 66 | Franco-Spanish War* | 106 | Italo-Turkish War |
| 27 | Scottish War* | 67 | War of the League of Augusburg* | 107 | World War I* |
| 28 | Spanish-Turkish War* | 68 | Second Northern War* | 108 | Russian Civil War* |
| 29 | First Huguenot War | 69 | War of the Spanish Succession* | 109 | Manchurian War |
| 30 | Austro-Turkish War* | 70 | Ottoman War | 110 | Italo-Ethiopian War |
| 31 | Spanish-Turkish War* | 71 | War of the Quadruple Alliance* | 111 | Sino-Japanese War |
| 32 | Austro-Turkish War* | 72 | British-Spanish War* | 112 | Russo-Japanese War* |
| 33 | Spanish-Potuguese War | 73 | War of the Polish Succession* | 113 | World War II* |
| 34 | Polish-Turkish War | 74 | Ottoman War | 114 | Russo-Finnish War |
| 35 | War of the Armada* | 75 | War of the Austrian Succession* | 115 | Korean War* |
| 36 | Austro-Polish War | 76 | Russo-Swedish War | 116 | Russo-Hungarian War |
| 37 | War of the Three Henries* | 77 | Seven Years' War* | 117 | Sinai War |
| 38 | Austro-Turkish War* | 78 | Russo-Turkish War | 118 | Sino-Indian War |
| 39 | Franco-Savoian War | 79 | Confederation of Bar | 119 | Vietnam War |
| 40 | Spanish-Turkish War* | 80 | War of the Bavarian Succession* | | |

## 2. Development of the size of wars over time.

I began by determining the size distribution of Great Power wars (with size defined as a fraction, 1495–1945).
The type of distribution shows similarities with a power law distribution. A power law – with 'size' defined as the number of battle casualties – has been previously identified (Richardson 1960).

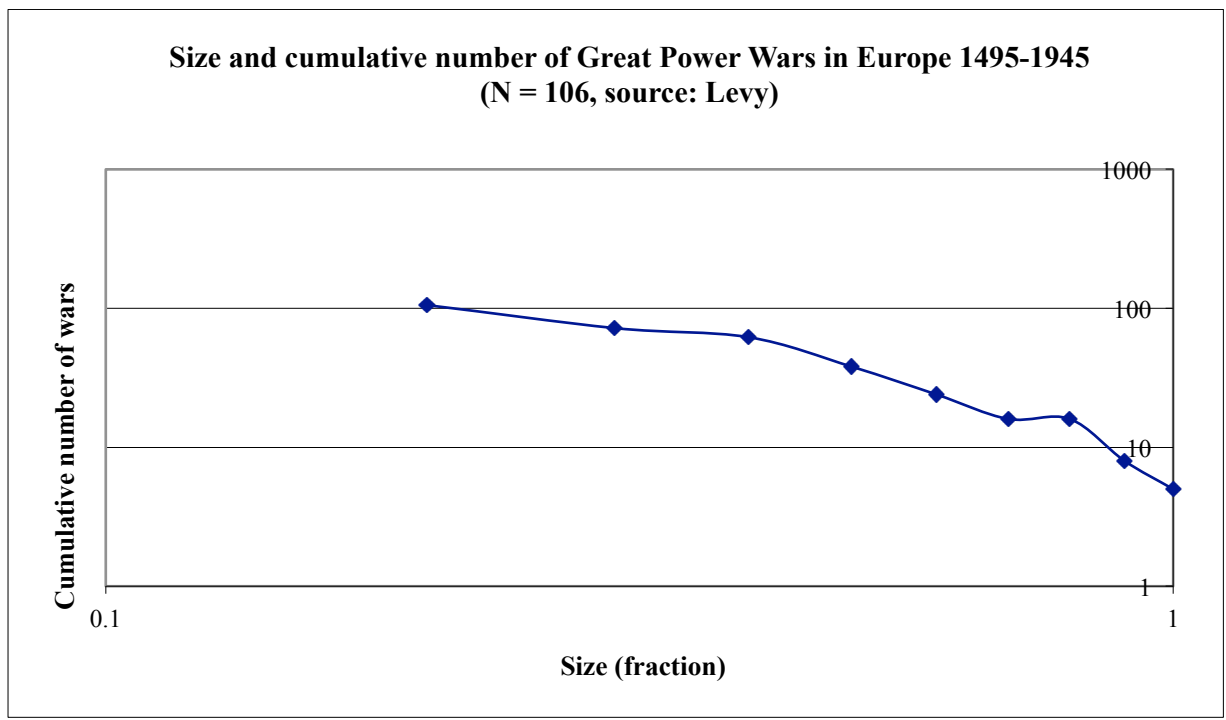

*Figure 1. Distribution of war size on a double logarithmic scale. The x-axis represents fraction, and the y-axis represents the number of wars.*

To gain a better understanding of the war dynamics of the International System, I examined how the fractions – the relative size – of Great Power wars have developed over time.
To obtain a more 'regular' graph of the 'fraction dynamics' of the International System, I repeatedly calculated the progressive mean of five consecutive war fractions. The value of the Great Power war corresponding with number '1' (see the *x*-axis), for example, is the mean of the fractions of the first five Great Power wars in Levy's dataset (Levy 1983, 88-91), the number '2' corresponds with war numbers 2 – 6, etc. The results of this analysis are shown in the figure below.
The thick line is a schematic, simplified illustration of the dynamic that can be identified.
In this figure, the abnormal war dynamics during the exceptional period are marked with a circle.

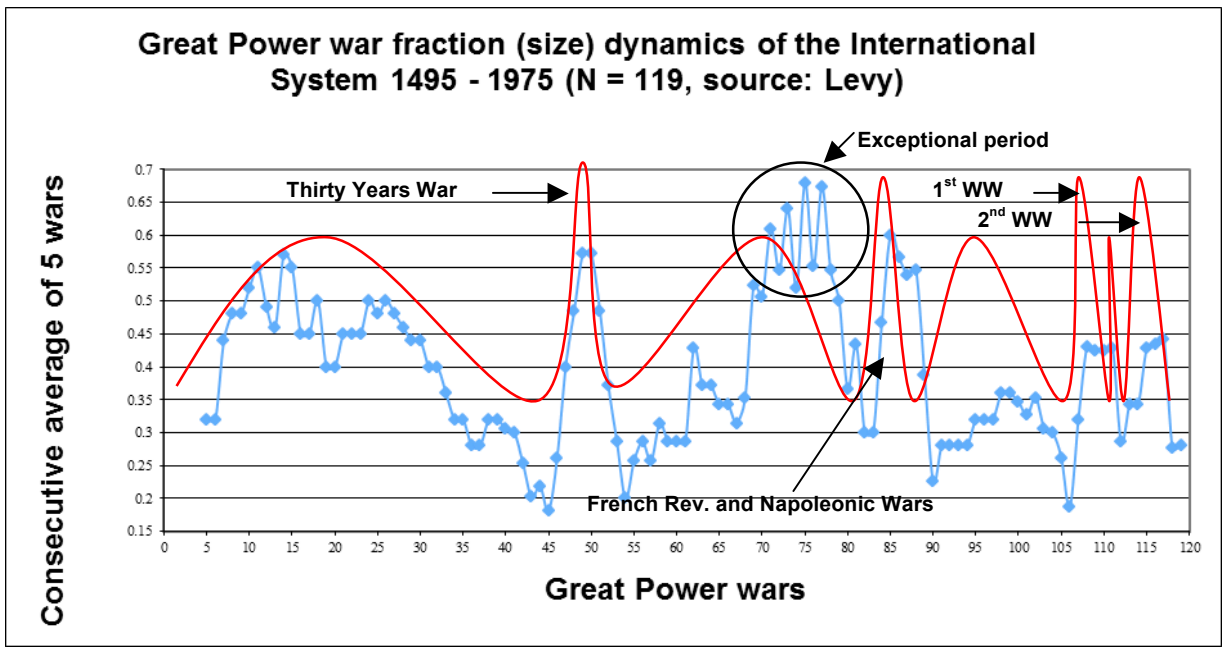

*Figure 2. Great Power war size dynamic, including several wars following the Second World War.*

### 3. Data used to identify chaotic dynamics of non-systemic wars.

| Orbits in phase state (size and intensity) of the International System | | | | | | |
|---|---|---|---|---|---|---|
| **Nr.** | **Shape** | **Start** | **End** | **Cycle** | **Direction** | **Duration** |
| 1 | Orbit (1) | 1495 | 1509 | 1 | R | 14 |
| 2 | Orbit (2) | 1511 | 1526 | 1 | L | 15 |
| 3 | Orbit (3) | 1521 | 1535 | 1 | L | 14 |
| 4 | Orbit (4) | 1532 | 1550 | 1 | L | 18 |
| 5 | Undefined | 1542 | 1556 | 1 | NA | 14 |
| 6 | Orbit (5) | 1556 | 1580 | 1 | R | 24 |
| 7 | Orbit (6) | 1576 | 1604 | 1 | R | 28 |
| 8 | Orbit (7) | 1587 | 1601 | 1 | L | 14 |
| 9 | Undefined | 1610 | 1621 | 1 | NA | 11 |
| 10 | Undefined | 1642 | 1668 | 2 | NA | 26 |
| 11 | Zigzag | 1657 | 1763 | 2 | NA | 106 |
| 12 | Orbit (8) | 1768 | 1790 | 2 | L | 22 |
| 13 | Orbit (9) | 1808 | 1849 | 3 | L | 41 |
| 14 | Undefined | 1849 | 1859 | 3 | NA | 10 |
| 15 | Orbit (10) | 1864 | 1912 | 3 | L | 48 |

*Table 1. Specifications of circular trajectories – orbits in the phase state – that can be identified.*

| Nr | Nr Levy | Nr Orbit | Direction | Intensity | Fraction | Start | End | I d 1000 |
|----|---------|----------|-----------|-----------|----------|-------|-----|----------|
| 1 | 1 | 1 | R | 119 | -0,60 | 1495 | 1497 | 0,119 |
| 2 | 2 | 1 | R | 45 | -0,20 | 1497 | 1498 | 0,045 |
| 3 | 3 | 1 | R | 60 | -0,20 | 1499 | 1503 | 0,06 |
| 4 | 4 | 1 | R | 29 | -0,20 | 1499 | 1500 | 0,029 |
| 5 | 5 | 1 | R | 269 | -0,40 | 1501 | 1504 | 0,269 |
| 6 | 6 | 1 | R | 145 | -0,60 | 1508 | 1509 | 0,145 |
| 7 | 7 | 2 | L | 261 | 0,80 | 1511 | 1514 | 0,261 |
| 8 | 8 | 2 | L | 343 | 0,40 | 1512 | 1519 | 0,343 |
| 9 | 9 | 2 | L | 57 | 0,20 | 1513 | 1515 | 0,057 |
| 10 | 10 | 2 | L | 43 | 0,60 | 1515 | 1515 | 0,043 |
| 11 | 11 | 2 | L | 420 | 0,75 | 1521 | 1526 | 0,42 |
| 12 | 12 | 3 | L | 958 | 0,50 | 1521 | 1531 | 0,958 |
| 13 | 13 | 3 | L | 41 | 0,25 | 1522 | 1523 | 0,041 |
| 14 | 14 | 3 | L | 249 | 0,75 | 1526 | 1529 | 0,249 |
| 15 | 15 | 3 | L | 384 | 0,50 | 1532 | 1535 | 0,384 |
| 16 | 16 | 4 | L | 55 | 0,25 | 1532 | 1534 | 0,055 |
| 17 | 17 | 4 | L | 438 | 0,50 | 1536 | 1538 | 0,438 |
| 18 | 18 | 4 | L | 1329 | 0,50 | 1537 | 1547 | 1,329 |
| 19 | 19 | 4 | L | 176 | 0,25 | 1542 | 1550 | 0,176 |
| 20 | 25 | 5 | R | 676 | -0,40 | 1556 | 1562 | 0,676 |
| 21 | 26 | 5 | R | 316 | -0,60 | 1556 | 1559 | 0,316 |
| 22 | 27 | 5 | R | 78 | -0,40 | 1559 | 1560 | 0,078 |
| 23 | 28 | 5 | R | 310 | -0,40 | 1559 | 1564 | 0,31 |
| 24 | 29 | 5 | R | 77 | -0,40 | 1562 | 1564 | 0,077 |
| 25 | 30 | 5 | R | 306 | -0,40 | 1565 | 1568 | 0,306 |
| 26 | 31 | 5 | R | 608 | -0,40 | 1569 | 1580 | 0,608 |
| 27 | 32 | 6 | R | 600 | -0,40 | 1576 | 1583 | 0,6 |
| 28 | 33 | 6 | R | 50 | -0,20 | 1579 | 1581 | 0,05 |
| 29 | 34 | 6 | R | 210 | -0,20 | 1583 | 1590 | 0,21 |
| 30 | 35 | 6 | R | 588 | -0,40 | 1585 | 1604 | 0,588 |
| 31 | 36 | 7 | L | 49 | 0,20 | 1587 | 1588 | 0,049 |
| 32 | 37 | 7 | L | 195 | 0,40 | 1589 | 1598 | 0,195 |
| 33 | 38 | 7 | L | 1086 | 0,40 | 1593 | 1606 | 1,086 |
| 34 | 39 | 7 | L | 24 | 0,20 | 1600 | 1601 | 0,024 |
| 35 | 78 | 8 | L | 127 | 0,17 | 1768 | 1774 | 0,127 |
| 36 | 79 | 8 | L | 149 | 0,17 | 1768 | 1772 | 0,149 |
| 37 | 80 | 8 | L | 3 | 0,33 | 1778 | 1779 | 0,003 |
| 38 | 82 | 8 | L | 1685 | 0,33 | 1787 | 1792 | 1,685 |
| 39 | 83 | 8 | L | 26 | 0,17 | 1788 | 1790 | 0,026 |
| 40 | 87 | 9 | L | 51 | 0,20 | 1808 | 1809 | 0,051 |
| 41 | 88 | 9 | L | 34 | 0,20 | 1812 | 1814 | 0,034 |
| 42 | 89 | 9 | L | 17 | 0,20 | 1815 | 1815 | 0,017 |
| 43 | 90 | 9 | L | 3 | 0,20 | 1823 | 1823 | 0,003 |
| 44 | 91 | 9 | L | 2 | 0,60 | 1827 | 1827 | 0,002 |
| 45 | 92 | 9 | L | 415 | 0,20 | 1828 | 1829 | 0,415 |
| 46 | 93 | 9 | L | 45 | 0,20 | 1848 | 1849 | 0,045 |
| 47 | 100 | 10 | L | 12 | 0,33 | 1864 | 1864 | 0,012 |
| 48 | 101 | 10 | L | 270 | 0,50 | 1866 | 1866 | 0,27 |
| 49 | 102 | 10 | L | 1415 | 0,33 | 1870 | 1871 | 1,415 |
| 50 | 103 | 10 | L | 935 | 0,17 | 1877 | 1878 | 0,935 |
| 51 | 106 | 10 | L | 45 | 0,13 | 1911 | 1912 | 0,045 |

*Table 2. An overview of the orbits that can be identified during the life span of the first, second, and third cycles (light, dark, and light blue, respectively). During the life span of the fourth cycle, there was only one war.*